\documentclass[reprint, amsmath,prx,longbibliography,amssymb, aps, superscriptaddress, floatfix]{revtex4-1}

\usepackage[utf8]{inputenc}
\usepackage{graphicx}
\usepackage{dcolumn}
\usepackage{bm}
\usepackage{braket}
\usepackage{color}
\usepackage{units}
\usepackage{dsfont}
\usepackage{float}
\usepackage{hyperref}
\usepackage{xcolor}

\begin{document}
\title{Measuring ultrafast time-bin qudits}

\author{Fr\'ed\'eric Bouchard}
\email{frederic.bouchard@nrc-cnrc.gc.ca}
\affiliation{National Research Council of Canada, 100 Sussex Drive, Ottawa, Ontario K1A 0R6, Canada}
\author{Kent Bonsma-Fisher}
\affiliation{National Research Council of Canada, 100 Sussex Drive, Ottawa, Ontario K1A 0R6, Canada}
\author{Khabat Heshami}
\affiliation{National Research Council of Canada, 100 Sussex Drive, Ottawa, Ontario K1A 0R6, Canada}
\affiliation{Department of Physics, University of Ottawa, Advanced Research Complex, 25 Templeton Street, Ottawa ON Canada, K1N 6N5}
\author{Philip J. Bustard}
\affiliation{National Research Council of Canada, 100 Sussex Drive, Ottawa, Ontario K1A 0R6, Canada}
\author{Duncan England}
\affiliation{National Research Council of Canada, 100 Sussex Drive, Ottawa, Ontario K1A 0R6, Canada}
\author{Benjamin Sussman}
\affiliation{National Research Council of Canada, 100 Sussex Drive, Ottawa, Ontario K1A 0R6, Canada}
\affiliation{Department of Physics, University of Ottawa, Advanced Research Complex, 25 Templeton Street, Ottawa ON Canada, K1N 6N5}

\begin{abstract}
Time-bin qudits have emerged as a promising encoding platform in many quantum photonic applications. However, the requirement for efficient single-shot measurement of time-bin qudits instead of reconstructive detection has restricted their widespread use in experiments. Here, we propose an efficient method to measure arbitrary superposition states of time-bin qudits and confirm it up to dimension 4. This method is based on encoding time bins at the picosecond time scale, also known as ultrafast time bins. By doing so, we enable the use of robust and phase-stable single spatial mode temporal interferometers to measure time-bin qudit in different measurement bases.
\end{abstract}

\maketitle

\section*{Introduction}

Photonic quantum information plays a central role in the greater quantum technology ecosystem~\cite{dowling2003quantum}. The ability to manipulate light at the quantum level has proven to be an invaluable tool for a variety of applications in quantum information processing~\cite{flamini2018photonic}, quantum communication~\cite{scarani:09}, and quantum sensing~\cite{pirandola2018advances}. The ability to efficiently measure photonic quantum states with high fidelity is at the heart of these quantum technologies. In particular, information can be encoded onto photons by various physical realizations. A non-exhaustive list of photonic degrees of freedom that can efficiently carry quantum information includes polarization~\cite{kwiat1995new}, time-bins~\cite{brendel1999pulsed,donohue2013coherent}, frequency-bins~\cite{lukens2017frequency}, position-bins~\cite{carine2020multi}, spatial modes~\cite{bouchard2018measuring}, temporal modes~\cite{brecht2015photon}, and electromagnetic field quadratures~\cite{braunstein2005quantum}. Polarization has often been the preferred degree of freedom to encode photonic quantum information due to its ease of generation, manipulation, and detection. However, other alternatives are being investigated to expand the potential use of photons in various quantum information tasks. 

Other photonic degrees of freedom, such as time and frequency, or position and momentum, may offer new, yet unexploited, advantages where polarization encoding is limited. One noticeable advantage of these degrees of freedom is their high-dimensional nature, i.e. \emph{qudits}, whereas polarization is bidimensional by nature, i.e. \emph{qubits}. High-dimensional quantum systems have been recognized as an essential resource for applications in quantum information processing~\cite{kues2017chip,wang2018multidimensional} and fundamental problems in quantum mechanics~\cite{malik2016multi,erhard2020advances}. In quantum communications, qudits are favourable in terms of information capacity~\cite{mower2013high} and noise tolerance~\cite{ecker2019overcoming}. More generally, quantum information processing requires a larger encoding space to perform simulation, computing, or sensing tasks~\cite{clements2016optimal}.

Although time-bin qubits have been widely employed in many quantum information tasks, time-bin qudits remain relatively unexplored. This can be credited to the lack of efficient, versatile and high-fidelity measurement schemes. Here, we propose a technique to efficiently measure time-bin qudits with high fidelity that is compatible with the requirements of most quantum information tasks. We perform a proof-of-principle quantum key distribution (QKD) experiment to demonstrate the benefit and convenience of our technique.

\begin{figure*}[t!]
	\centering
		\includegraphics[width=0.96\textwidth]{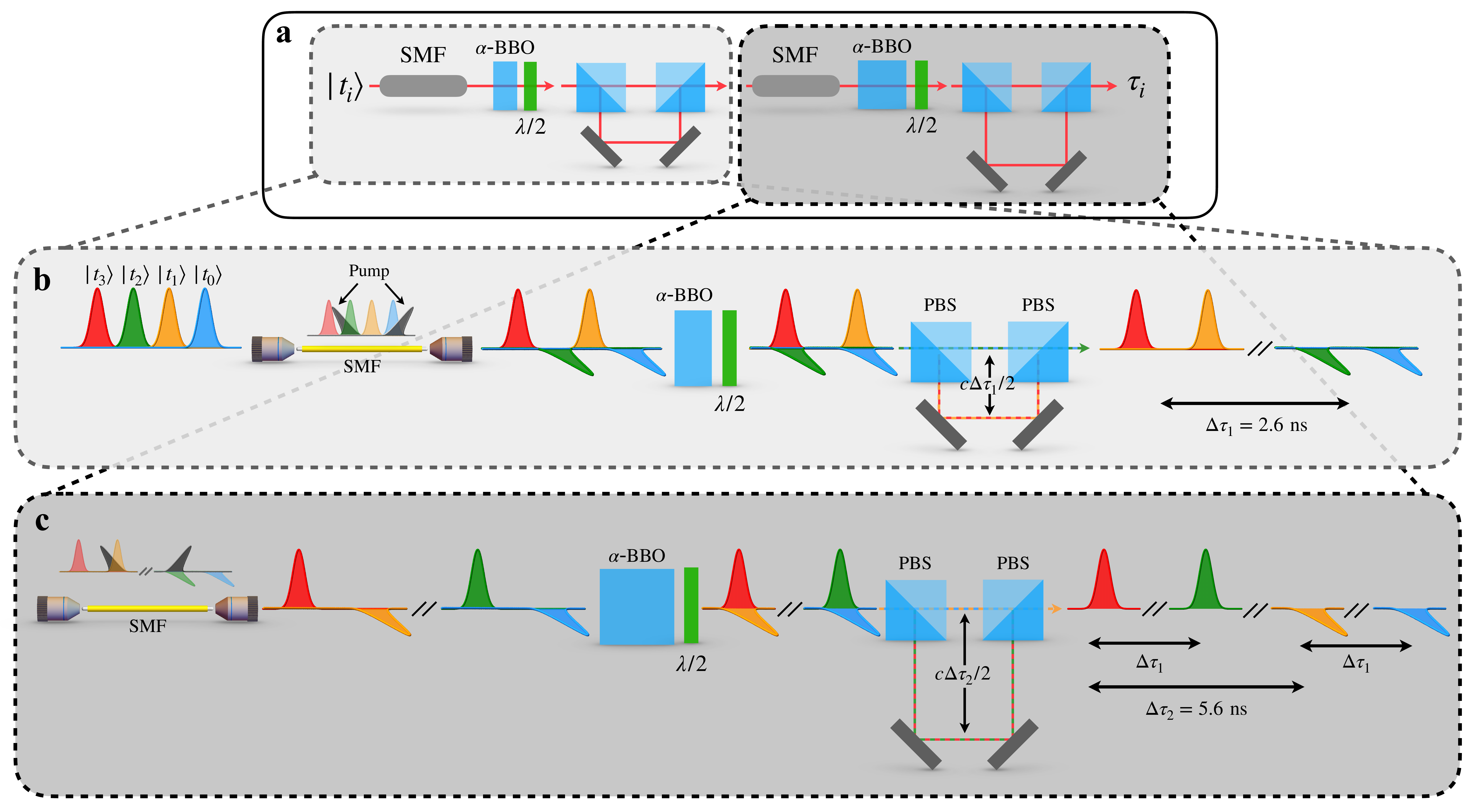}
	\caption{\textbf{Measuring time-bin qudits in the computational basis}. Simplified experimental setup for the measurement of computational time-bin qudits in dimension $d=4$. \textbf{(a)} Computational time-bin qudits, $|t_i\rangle$, are prepared and sent to a 14-cm single mode fiber (SMF), acting as the nonlinear medium, followed by a 5-mm $\alpha$-BBO crystal and a polarization time delay. As a second stage, the signal is sent to a second SMF followed by a 10-mm $\alpha$-BBO crystal and another polarization time delay. \textbf{(b)} A strong pump pulse at a center wavelength $\lambda_\mathrm{pump}$ is made to overlap with the time-bins $t_0$ and $t_2$. The polarization of the time bins at time $t_1$ and $t_3$ are left unchanged while the polarization of the time bins at time $t_0$ and $t_2$ are rotated by $\pi/2$. The $\alpha$-BBO crystal brings the pulses at time bins $t_1$ and $t_3$ to time bins $t_0$ and $t_2$. The polarization time delay with a path difference of 80~cm between each arm, corresponding to a temporal shift of $\Delta \tau_1=2.6$~ns, is used to temporally separate each polarization to different nanosecond scale time bins. \textbf{(c)} The signal is then sent in the second SMF where strong pump pulses are made to overlap with the quantum signal. Specific time bins will experience polarization rotation due to the strong pump pulse inside the nonlinear medium. Finally, the $\alpha$-BBO crystal and the polarization time delay with a path difference of 170~cm between each arm, corresponding to a temporal shift of $\Delta \tau_2=5.6$~ns, are used to temporally separate each polarization to additional nanosecond scale time bins. The final state can then be directly measured with a single photon detector since the computational qudits are now encoded in time bins with nanosecond temporal separation, which is larger than the timing jitter of the detector.} 
	\label{fig:setupcomp}
\end{figure*}

The ease by which time-bin states of photons can be measured in their computational basis is only matched by how difficult their measurements are in superposition bases. A fast single-photon detector can directly detect the time-of-arrival of photons, thus directly measuring the photon's state in the computational basis, i.e. ${|t_n\rangle \in \{ |t_0\rangle, |t_1\rangle, ..., |t_{d-1}\rangle \}}$, where $d$ is the dimension of the encoding space. Time-bin qudits can also be prepared in mutually unbiased basis (MUB) elements made from superpositions of time bins. In dimension $d$, a MUB example is given by,
\begin{eqnarray}
|\varphi_n\rangle = \frac{1}{\sqrt{d}} \sum_{m=0}^{d-1} \exp \left[ \frac{2 \pi i n m}{d} \right] |t_m\rangle,
\end{eqnarray}
for ${n=0,1,...,d-1}$. Here, we consider the following MUB in dimension 4; $|\phi_0\rangle=(|t_0\rangle+|t_1\rangle+|t_2\rangle+|t_3\rangle)/2$, $|\phi_1\rangle=(|t_0\rangle-|t_1\rangle+|t_2\rangle-|t_3\rangle)/2$, $|\phi_2\rangle=(|t_0\rangle+|t_1\rangle-|t_2\rangle-|t_3\rangle)/2$, and $|\phi_3\rangle=(|t_0\rangle-|t_1\rangle-|t_2\rangle+|t_3\rangle)/2$. In this MUB, an interferometric measurement among the time-bins is generally required. This is typically achieved by employing an imbalanced, or time-delayed, interferometer~\cite{marcikic2002time,brougham2013security}. In this configuration, the path difference between both arms of the interferometer is made to correspond to the time separation among time bins. By doing so, one can obtain information about the relative phase of specific time-bin states, but one generally needs to cascade time-delayed interferometers with varying path differences to obtain a measurement outcome on the overall state encoded from the MUB elements~\cite{islam2017provably,ikuta2017implementation}. With passive elements, this approach is inefficient since the time shifting occurring in the time-delayed interferometer also leads to non-interfering events that do not result in useful phase information. The fixed delays of such passive setups must be interferometrically stable with respect to the distinct time bins permitted by detector jitter and pulse duration, often raising the need for active phase stabilization.

\begin{figure*}[t!]
	\centering
		\includegraphics[width=0.96\textwidth]{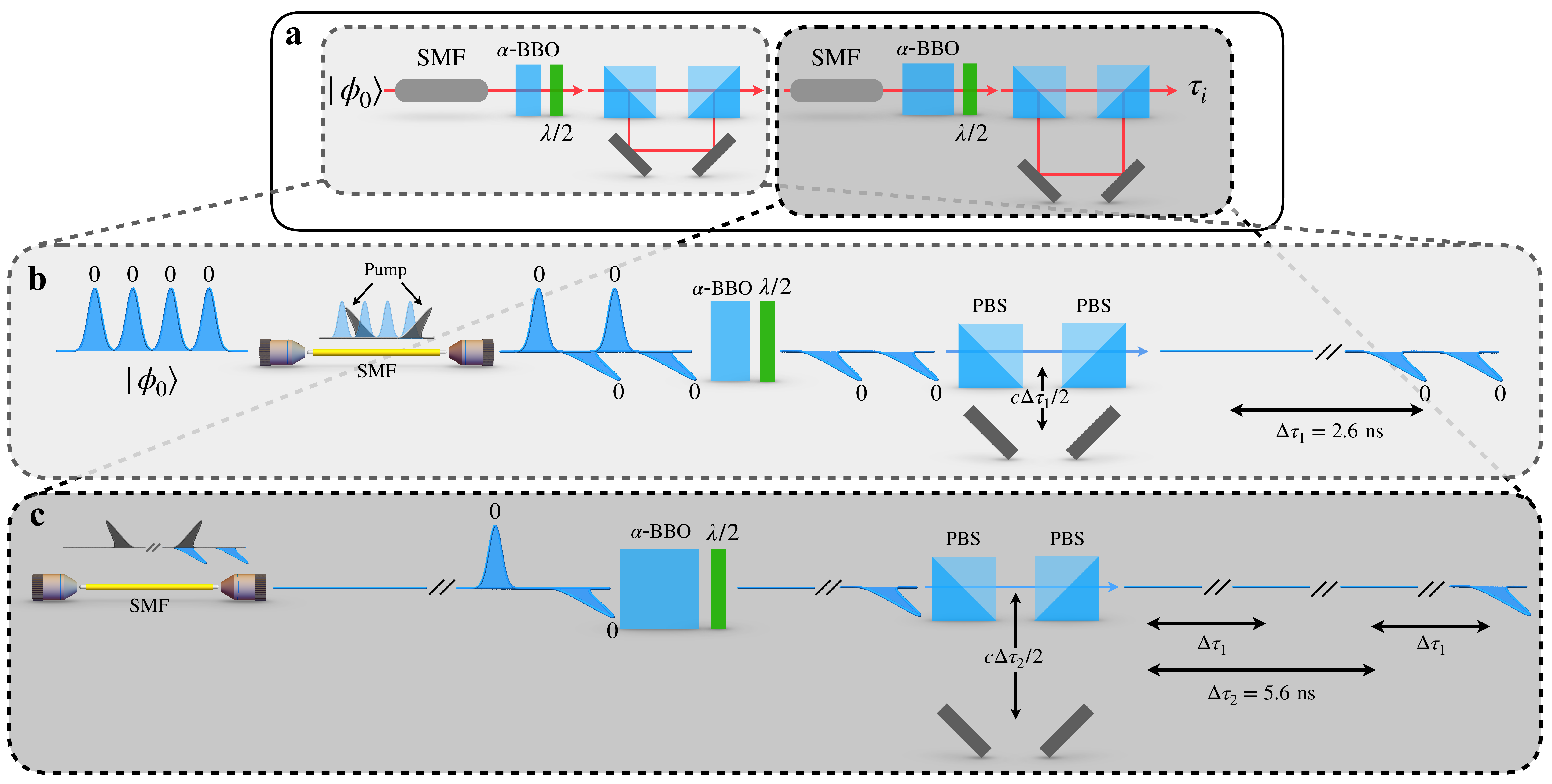}
	\caption{\textbf{Measuring time-bin qudits in the superposition basis}. Simplified experimental setup for the measurement of superposition time-bin qudits in dimension $d=4$. \textbf{(a)} The superposition time-bin qudit, $|\phi_0\rangle$, is prepared and sent to a 14-cm single mode fiber (SMF), acting as the nonlinear medium, followed by a 5-mm $\alpha$-BBO crystal, a half-wave plate ($\lambda/2$) and a polarization time delay. As a second stage, the signal is sent to a second SMF followed by a 10-mm $\alpha$-BBO crystal, a half-wave plate and another polarization time delay. \textbf{(b)} A strong pump pulse at a center wavelength $\lambda_\mathrm{pump}$ is made to overlap with the time-bins $t_0$ and $t_2$. The polarization of the time bins at time $t_1$ and $t_3$ are left unchanged while the polarization of the time bins at time $t_0$ and $t_2$ are rotated by $\pi/2$.  A combination of an $\alpha$-BBO crystal and a half-wave plate, $\lambda/2$, brings the pulses at time bins $t_1$ and $t_3$ to interfere with time bins at $t_0$ and $t_2$. The polarization time delay with a path difference of 80~cm between each arm, corresponding to a temporal shift of $\Delta \tau_1=2.6$~ns, is used to temporally separate each polarization to different nanosecond scale time bins. \textbf{(c)} The signal is then sent in the second SMF where strong pump pulses are made to overlap with the quantum signal. Specific time bins will experience polarization rotation due to the strong pump pulse inside the nonlinear medium. Finally, the $\alpha$-BBO crystal, half-wave plate and polarization time delay with a path difference of 170~cm between each arm, corresponding to a temporal shift of $\Delta \tau_2=5.6$~ns, are used to temporally separate each polarization to additional nanosecond scale time bins. The final state can then be directly measured with a single photon detector since the superposition qudits are now encoded in time bins with nanosecond temporal separation, which is larger than the timing jitter of the detector. See Fig.~\ref{fig:setup2}-\ref{fig:setup4} for the evolution of the states $|\phi_1\rangle$, $|\phi_2\rangle$, and $|\phi_3\rangle$ through the experimental setup.}
	\label{fig:setup}
\end{figure*}

To overcome the inefficiency of passive interferometric setups, active elements can be introduced to avoid non-interfering events. This can be achieved with fast electro-optic modulators or optical nonlinearity such as cross-phase modulation based on the optical Kerr effect in the presence of a strong pulsed laser~\cite{nowierski2016tomographic,kupchak2017time,kupchak2019terahertz}. However, in order to achieve an efficient single-shot measurement of a time-bin superposition state, the appropriate combination of active optical elements and time-delayed interferometers must be realized. Recently, we demonstrated the use of ultrafast time-bins, where time bin qubits are encoded onto picosecond-duration pulses~\cite{bouchard2022quantum}. One of the main advantages of operating at such a time scale is the notable reduction of the time separation among time bins. This has the direct effect of reducing the path difference of time-delayed interferometers to only a few hundred micrometers, thus offering the potential for compact time-delayed interferometers with intrinsic passive phase stability over long periods of time. Finally, other configurations based on fast optical modulation and spectral phase modulation have been proposed and investigated numerically~\cite{lukens2018reconfigurable,ashby2020temporal}, but experimental demonstrations have yet to be reported.

\begin{figure*}[t!]
	\centering
		\includegraphics[width=0.9\textwidth]{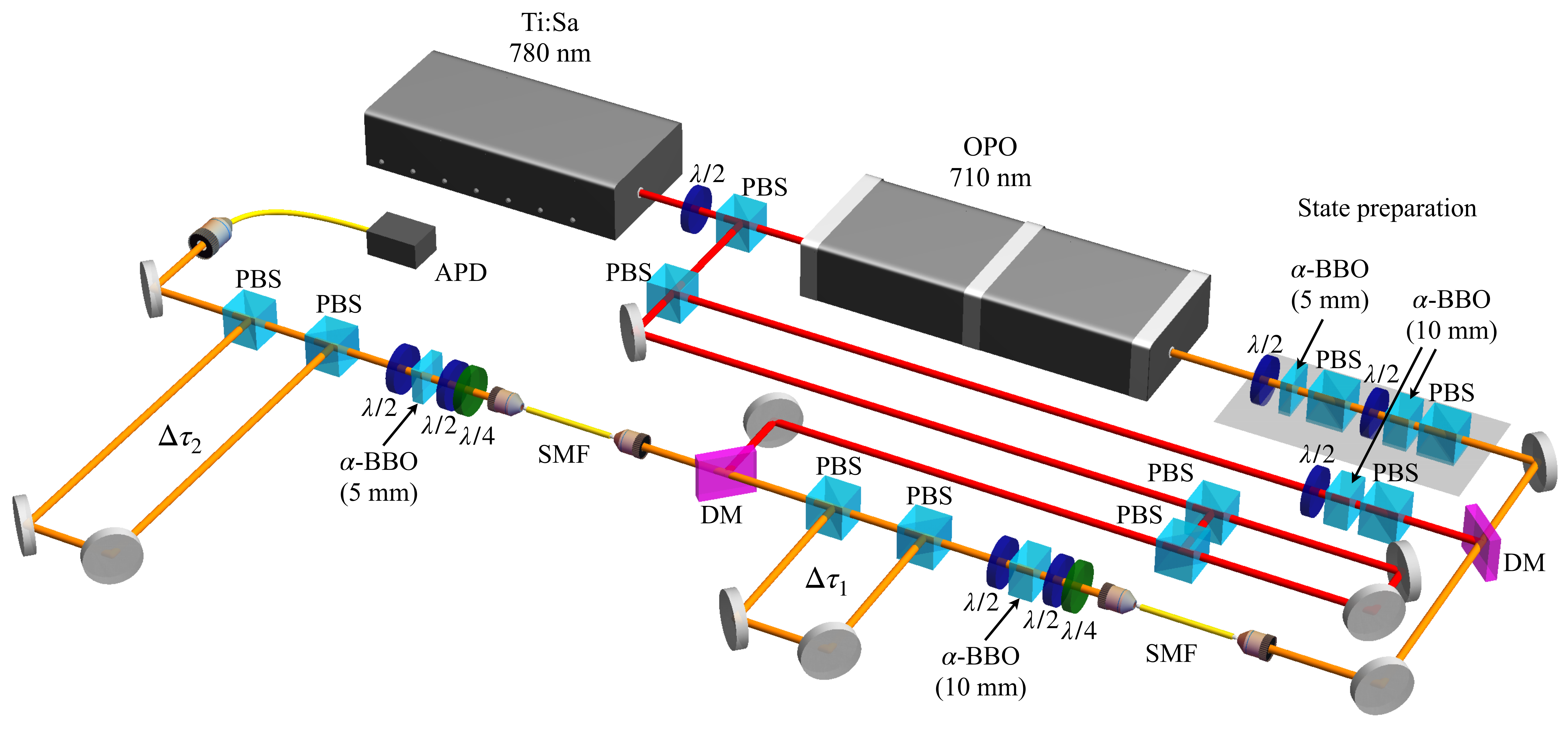}
	\caption{\textbf{Experimental setup}. Experimental setup demonstrating the measurement of ultrafast time-bin qudits in dimension $d=4$. Ti:Sa: Titanium Sapphire laser; $\lambda$/2: Half-wave plate; $\lambda$/4: Quarter-wave plate; PBS: Polarizing beam splitter; OPO: Optical parametric oscillator; ND: Neutral density filter; BS: beam splitter; DM: Dichroic mirror; SMF: Single mode fibre; APD: Avalanche photodiode detectors.}
	\label{fig:detailedsetup}
\end{figure*}

\section*{Experiment}

Here, we propose and experimentally demonstrate a technique to measure time-bin qudits in a single-shot setting. The practicality of our measurement scheme is then tested in a proof-of-principle high-dimensional quantum communication demonstration. Ultrafast time-bin qudits in dimension $d=4$ are realized with a bin separation of $\tau = 2.25$~ps. The small time difference among time-bins enables us to use compact time-delayed interferometers. In particular, we take advantage of the difference in group index experienced by orthogonal linear polarization states in a birefringent crystal to achieve a collinear time-delayed interferometer. An intrinsic phase stability is realized by having both \emph{arms} of the time-delayed interferometers in a single spatial mode. At a signal wavelength of $\lambda_\mathrm{signal}=720\mathrm{nm}$, a 5-mm-long $\alpha$-barium borate (BBO) crystal can give rise to a group delay of $\tau=2.25$~ps between orthogonal linearly polarized pulses.

In the computational basis, time-bin states are measured using an ultrafast polarization switch (UPS), where the polarization of the time-bin that is measured is rotated by an angle of $90^\circ$.  The UPS is achieved via cross-phase modulation in a 14-cm SM600 single-mode fiber (SMF) by applying a strong pump pulse. We note that the pulse energies required to induce a nonlinear phase shift of $\Delta \phi = \pi $, in our experiment, are well below the point of optical damage in the fibre. The switching efficiency $\eta$ of the UPS is given by~\cite{agrawal2000nonlinear}
\begin{eqnarray}
\eta=\sin^2 \left(2 \theta \right) \sin^2 \left( \frac{\Delta \phi}{2} \right),
\end{eqnarray}
where $\theta$ is the angle between the polarization of the pump and the signal pulses, $\Delta \phi = 8 \pi n_2 L_\mathrm{eff} I_\mathrm{pump}/3 \lambda_\mathrm{signal}$ is the nonlinear phase shift induced by the pump pulse in the SMF, $n_2$ is the nonlinear refractive index of the SMF, $L_\mathrm{eff}$ is the effective length of the nonlinear medium, and $I_\mathrm{pump}$ is the intensity of the pump pulse. To achieve maximal switching efficiency, the pump pulse is polarized at an angle of $\theta=\pi/4$ with respect to the polarization of the signal pulses and the pump intensity is adjusted to achieve $\Delta \phi = \pi$. A single shot measurement of the time-bin qudits is realized by cascading two UPS, see Fig.~\ref{fig:setupcomp}. Using the same experimental setup with additional half-wave plates, time-bin qudits prepared in the superposition basis can also be measured, see Fig.~\ref{fig:setup} and Appendix A. In order to switch between measurement basis, the half-wave plate is simply rotated from a $0^\circ$ angle for the computational basis to a $22.5^\circ$ angle for the superposition basis. We note that the pump pulse may induce an additional global phase onto the rotated time-bin signal. In our experiment, this additional phase is corrected by using a sequence of wave plates.

Weak coherent pulses are generated at a wavelength of $\lambda_\mathrm{sig}=710$~nm by pumping an optical parametric oscillator (OPO) with a Ti:sapphire laser at a wavelength and a repetition rate of $\lambda_\mathrm{pump}=780$~nm and $f_\mathrm{rep}=80$~MHz, respectively, see Fig.~\ref{fig:detailedsetup}.  The signal and pump wavelengths are selected in order to limit the amount of parasitic noise generated by the strong pump pulses. In particular, nonlinear processes, such as self-phase modulation and two-photon absorption, may create noise photons covering the spectral range of interest for the quantum signals~\cite{england2021perspectives}. The signal and pump pulses are spectrally filtered to a bandwidth of $\Delta \lambda_\mathrm{signal}=5.0$~nm and $\Delta \lambda_\mathrm{pump}=5.8$~nm, corresponding to transform-limited pulse durations of 0.15~ps and 0.15~ps, respectively. The time-bin qudit states are prepared using a sequence of a half-wave plate (HWP), a 5-mm $\alpha$-BBO crystal, a polarizing beam splitter (PBS), a HWP, a 10-mm $\alpha$-BBO crystal, and another PBS. The $\alpha$-BBO crystals are set at a $45^\circ$ angle from the PBS and the rotation angles of both HWPs then determine the generated time-bin qudit states. In the computational basis, the states $|t_0\rangle$, $|t_1\rangle$, $|t_2\rangle$, and $|t_3\rangle$ are achieved by setting the HWP angles to $22.5^\circ$/$22.5^\circ$, $-22.5^\circ$/$22.5^\circ$, $22.5^\circ$/$-22.5^\circ$, and $-22.5^\circ$/$-22.5^\circ$, respectively. Similarly, in the superposition basis, the states $|\phi_0\rangle$, $|\phi_1\rangle$, $|\phi_2\rangle$, and $|\phi_3\rangle$ are achieved by setting the HWP angles to $0^\circ$/$0^\circ$, $45^\circ$/$0^\circ$, $0^\circ$/$45^\circ$, and $45^\circ$/$45^\circ$, respectively. We note that using this method, an overall state generation efficiency of 25~\% is achieved due to post-selection with the PBS. This does not constitute a limitation since the weak coherent pulses must be attenuated to single photon levels regardless. However, efficient single-photon state generation can also be achieved using active switching with the UPS. Finally, the weak coherent pulses are attenuated to the single photon level. In our experiment, a mean photon number of $\mu=0.14$ per pulse is measured. In a real quantum communication setting, this parameter is adjusted to maximize the overall secret key rate of the system. Nevertheless, our experimental value is consistent with mean photon numbers reported in experimental demonstrations of the decoy-state BB84 protocol~\cite{hwang2003quantum}.

\begin{figure*}[t!]
	\centering
		\includegraphics[width=0.98\textwidth]{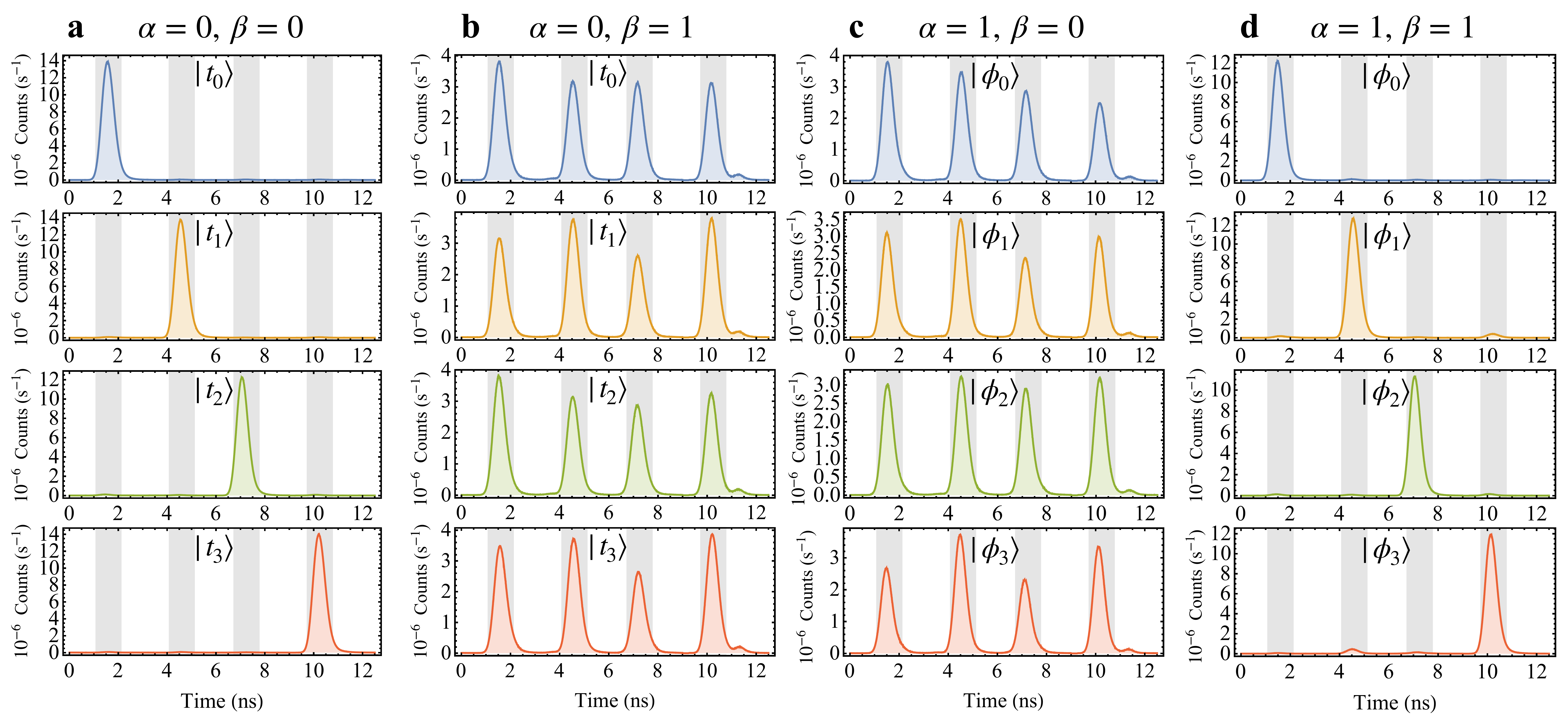}
	\caption{\textbf{Time-bin qudit measurements}. Detection events, $N_{i,j}^{(\alpha,\beta)}$ for (a) $\alpha=0$, $\beta=0$, (b) $\alpha=0$, $\beta=1$, (c) $\alpha=1$, $\beta=0$, (d) $\alpha=1$, $\beta=1$. The shaded gray area represents the 1~ns temporal windows corresponding to each time-bin state. The number of detection events shown corresponds to an integration time of 1~s. The cases for which {$\alpha=0$} and {$\alpha=1$} correspond to Alice preparing the input states in the computational and superposition bases, respectively. The cases for which {$\beta=0$} and {$\beta=1$} correspond to Bob measuring the incoming states in the computational and superposition bases, respectively.}
	\label{fig:resultsN}
\end{figure*}

At the measurement stage, the input states are sent to the first SMF with a coupling efficiency of 80~\%. Two synchronized pump pulses are also coupled in the same SMF with a total pulse energy of 4.7~nJ for both pulses and coupling efficiency of 70~\%. The pumps are prepared in order to overlap with the signal time-bins $|t_0\rangle$ and $|t_2\rangle$. This is done by using a 10-mm $\alpha$-BBO crystal on the pump beam. Here, we note that the pump pulses will have orthogonal polarization which will result in the rotated signal to acquire an additional $\pi$ phase from the UPS. This extra phase (not represented in Fig.~\ref{fig:setup}) is simply compensated for in the state preparation. On exiting the SMF, the signal propagates through a 5-mm $\alpha$-BBO crystal which has the effect of recombining time bins $t_1$ with $t_0$ and $t_3$ with $t_2$. In the superposition basis, this leads to the interference of adjacent time bins, where the relative phase information is mapped onto polarization. The signal beam is then sent to a polarization time delay, with a path difference of 80~cm, used to achieve a fixed temporal shift of $\Delta \tau_1=2.6$~ns, in order to temporally separate the time bins with horizontal and vertical polarization to different nanosecond scale time bins. We note that the polarization time delay is only used to map polarization to distinct nanosecond time bins and does not require phase stability since no interference occurs at the output ports. As a second stage, the signal is coupled to a second SMF with a coupling efficiency of 76~\%. Two orthogonally polarized pump beams prepared to overlap with the signal pulses at the appropriate time bins are also coupled to the second SMF with input pulse energy of 2.7~nJ and 3.2~nJ, and coupling efficiencies of 64~\% and 62~\%, respectively. Finally, a 10-mm $\alpha$-BBO crystal recombines the signal time bin $t_2$ with $t_0$, where, once again, in the superposition basis, the relative phase of the time bins is mapped on the polarization of the final $t_0$ time bin. A final polarization time delay with a path difference of 170~cm, corresponding to a temporal shift of $\Delta \tau_2=5.6$~ns, in order to temporally separates horizontal and vertical polarization to different nanosecond scale time bins. In both measurement bases, different input states will ultimately be mapped to different nanosecond scale time-bins that can be measured with a standard avalanche photodiode detector (APD).

\begin{figure}[t!]
	\centering
		\includegraphics[width=0.48\textwidth]{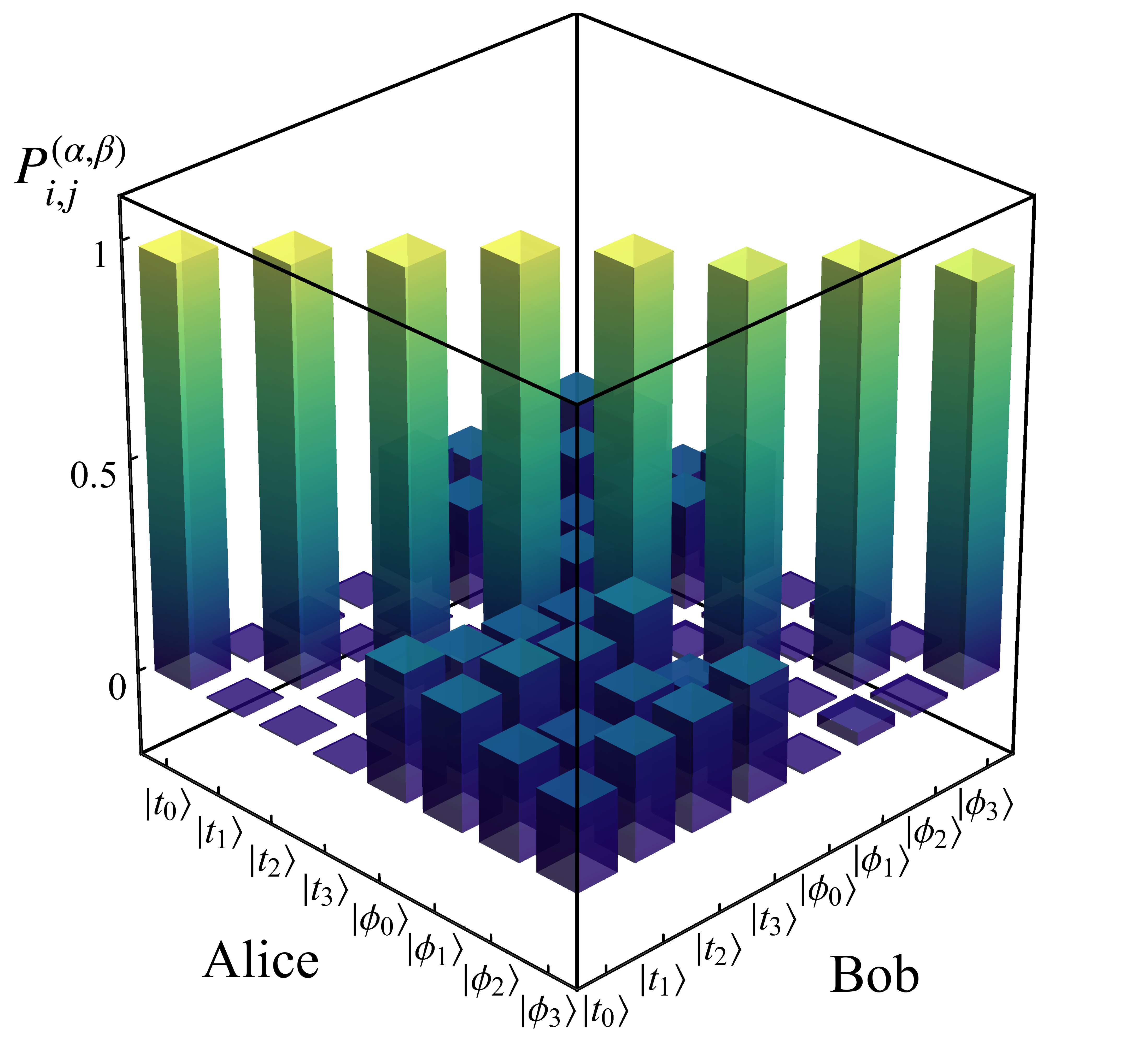}
	\caption{\textbf{Probability of detection measurements}. From the experimental detection events, the probability of detection is obtained.}
	\label{fig:resultsP}
\end{figure}

\section*{Results}

We analyze the performance of our measurement scheme in the context of QKD where a sender, typically referred to as \emph{Alice}, prepares the time-bin states and sends them to a receiver, typically referred to as \emph{Bob}. Bob's measurement apparatus will be characterized experimentally by measuring detection events $N_{i,j}^{(\alpha,\beta)}$ corresponding to Alice preparing a state $|\psi_i^{(\alpha)}\rangle$ and Bob measuring a state $|\psi_j^{(\beta)}\rangle$. The computational basis corresponds to ${\alpha,\beta=0}$ and the superposition basis to ${\alpha,\beta=1}$. From detection events, it is possible to obtain the probability of detection of corresponding states, i.e. ${P_{i,j}^{(\alpha,\beta)}= \left|\langle \psi_j^{(\beta)} | \psi_i^{(\alpha)} \rangle \right|^2=N_{i,j}^{(\alpha,\beta)}/\sum_{k=0}^3 N_{i,k}^{(\alpha,\beta)}}$, where ${|\psi_i^{(0)}\rangle=|t_i\rangle}$ and ${|\psi_i^{(1)}\rangle=|\phi_i\rangle}$ for $i \in \{0,1,2,3\}$. The $\alpha$ and $\beta$ bases correspond to Alice's and Bob's generation and measurement bases, respectively. Finally, we define the state fidelity, ${F_i^{(\alpha)}=P_{i,i}^{(\alpha,\alpha)}}$ in terms of probability of detections. In the BB84 QKD protocol, an important figure of merit is the quantum bit error rate (QBER), $Q$, and can be experimentally obtained from the average state fidelity, i.e. $Q=1-\frac{1}{2d}\sum_{i=0}^{(d-1)} \left(F_i^{(0)}+F_i^{(1)} \right)$.

The detection events, $N_{i,j}^{(\alpha,\beta)}$, are shown in Fig.~\ref{fig:resultsN}. The single-photon signal measured by the APD is shown as a function of time at the nanosecond scale. We assign temporal windows of 1~ns for each time-bin qudit state over which the signal is integrated. From the experimental detection event, we show the corresponding probabilities of detection, see Fig.~\ref{fig:resultsP}. In the computational basis, the state fidelities are given by ${F_0^{(0)} = 98.7~\%}$, ${F_1^{(0)} = 98.4~\%}$, ${F_2^{(0)} = 97.8~\%}$, and ${F_3^{(0)} = 98.6~\%}$. In the superposition basis, the state fidelities are given by ${F_0^{(1)} = 97.8~\%}$, ${F_1^{(1)} = 94.8~\%}$, ${F_2^{(1)} = 96.5~\%}$, and ${F_3^{(1)} = 94.5~\%}$. In a BB84 QKD protocol, the corresponding QBER is given by ${Q=2.8~\%}$. The performance of a quantum communication system is given by the secret key rate, $R$. In particular, the secret key rate for high-dimensional BB84 QKD protocol is given by~\cite{sheridan2010security},
\begin{eqnarray}
R^{(d)}(Q)=\log_2(d) - 2 h^{(d)}(Q),
\end{eqnarray}
where ${h^{(d)}(x):= -x \log_2 (x/(d-1)) - (1-x) \log_2 (1-x)  }$ is the $d$-dimensional Shannon entropy. From the measured QBER, we obtain a secret key rate of $R=1.54$~bits per sifted photon, which is beyond the $R^{(2)}(0)=1$~bit point where qudits can increase the information capacity per detected photon of quantum communication system. We note that a diminished secret key rate is expected when considering the security analysis of the decoy state protocol~\cite{ma2005practical}. Nevertheless, with sufficiently low noise and loss, an improvement in secret key rate is still expected from a high-dimensional protocol using our measurement scheme.

\section*{Discussion and outlook}

High-dimensional quantum communication schemes have been investigated in many settings~\cite{mower2013high,sit2017high,ding2017high,islam2017provably,bouchard2018experimental}, with potential applications with noise resilience and larger rates. For the case of noise resilience, qudits may have an advantage over their qubit counterpart, given a certain set of channel conditions on noise characteristics of the generation and detection systems~\cite{zhu2021high}. For a different type of channel, i.e. short communication links with low loss, where the detectors are operated close to their saturation point, one may reach the limit of the number of photons that can be detected per second. However, the users in quantum communication systems are interested in the secret key rate, which can be increased by encoding more than one bit of secret key per photon. By doing so, qudits can drastically improve the performance of quantum communication systems. We note here that our ultrafast time-bin qudit platform supports both avenues of operating at large rates and tolerating larger amounts of noise. By encoding time-bins in the ultrafast regime, we are dealing with pulses that have a minimum time-bandwidth product, also known as Fourier-transform-limited, which are known to occupy a single spectrotemporal mode. By appropriately filtering the incoming signal, ultimate noise tolerance can be achieved~\cite{bouchard2021achieving}. Thus our scheme enables the full potential promised by qudits in quantum communications.

We limited our demonstration to the case of dimension $d=4$, due to the limited laser power available; however, the scheme may readily be extended to larger dimensions with sufficient resources. For the case of $d=2^n$, where ${n=1,2,3,...}$, one can cascade additional stages comprised of an SMF, an $\alpha$-BBO of appropriate size, and a polarization time delay with appropriate path difference. Additional pump pulses need to be prepared and overlapped with the correct signal time bins, but the excellent interferometric stability would still be maintained by $\alpha$-BBO crystals. We also note that dimensions that are not powers of 2 can be achieved with a slightly different and less compact experimental setup. With enough UPS and birefringent crystals, any arbitrary time-bin qudit states can be measured. The final separation of the measurement bins was dictated by the timing jitter of our APD detectors. Superconducting nanowire single photon detectors (SNSPDs) offer superior temporal resolution and high quantum efficiency. Combined with our active detection scheme, in principle the measurement bins could be much more closely spaced. This would reduce the size of the polarization time delays and increase the maximum possible bandwidth of the system.

In conclusion, we have experimentally demonstrated an efficient method to measure time-bin qudit states of single photons in dimension 4. This measurement is carried out in a single shot and not through a sequence of reconstructive measurements. Our method can be directly implemented and applied in several quantum photonic applications such as quantum communication, e.g. QKD. Moreover, we establish a robust and phase-stable platform to manipulate multi-dimensional temporal states of photons, a key ingredient in quantum information processing systems. By doing so, we provide a pathway to achieve temporal information processing using ultrafast time-bin states of light.

\section*{Acknowledgments}
This work is supported by the High Throughput Secure Networks Challenge Program at the National Research Council of Canada (NRC), the Natural Sciences and Engineering Research Council of Canada, and the University of Ottawa-NRC Joint Centre for Extreme Photonics. We thank Kate Fenwick, Rune Lausten, Denis Guay, and Doug Moffatt for support and insightful discussions.

\section*{Appendix A: Measurement of time-bin qudits in the superposition basis}

\begin{figure*}[t!]
	\centering
		\includegraphics[width=0.92\textwidth]{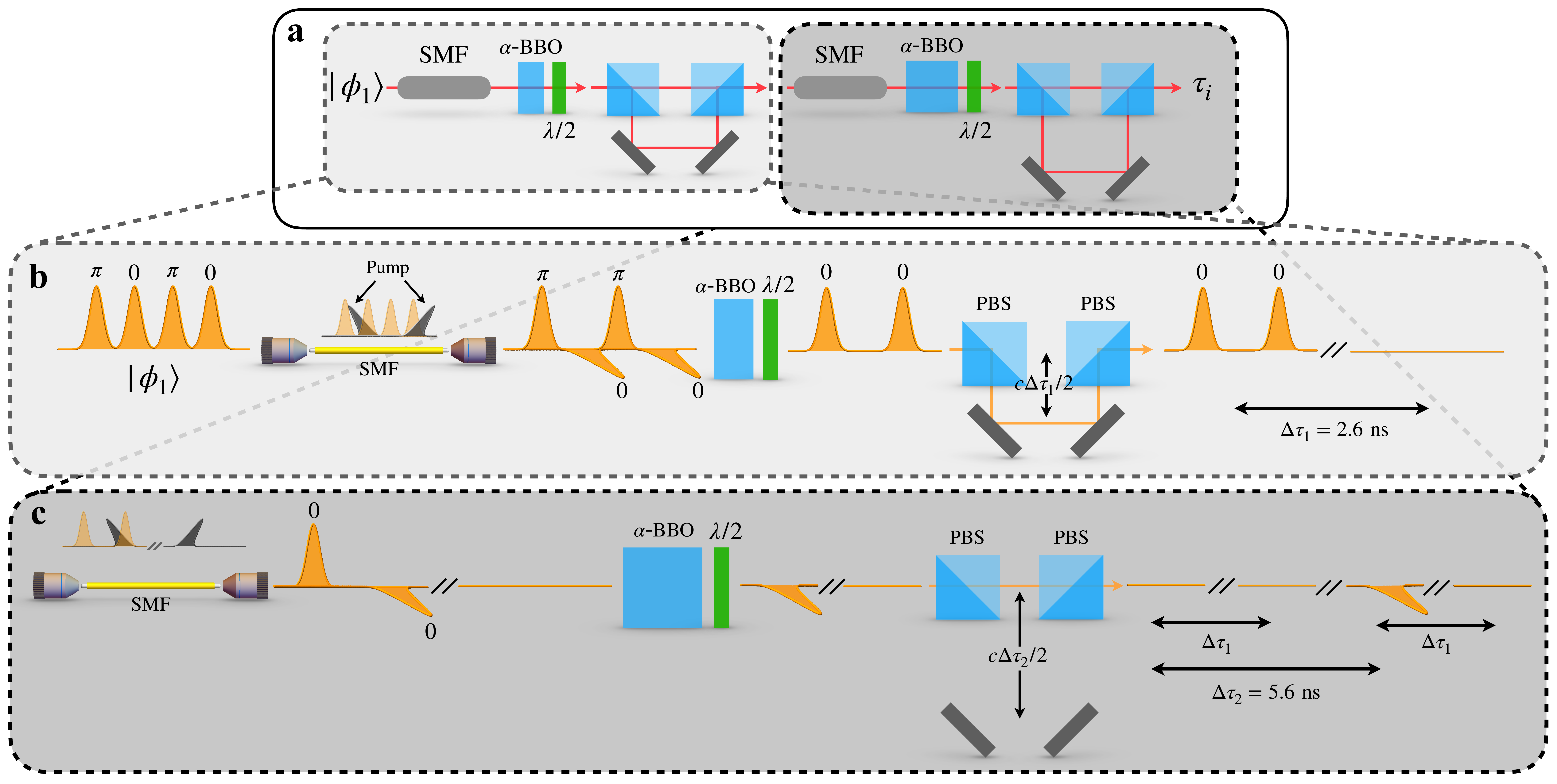}
	\caption{\textbf{Measuring time-bin qudits in the superposition basis with input state $|\phi_1\rangle$}.  \textbf{(a)} Simplified experimental setup for the measurement of superposition time-bin qudits in dimension $d=4$ for input state $|\phi_1\rangle$. \textbf{(b)} The first stage consists of an SMF, a 5-mm $\alpha$-BBO crystal, and a polarization time delay with a path difference of $\sim80$~cm. \textbf{(c)} The second stage consists of an SMF, a 10-mm $\alpha$-BBO crystal, and a polarization time delay with a path difference of $\sim170$~cm.}
	\label{fig:setup2}
\end{figure*}

\begin{figure*}[t!]
	\centering
		\includegraphics[width=0.92\textwidth]{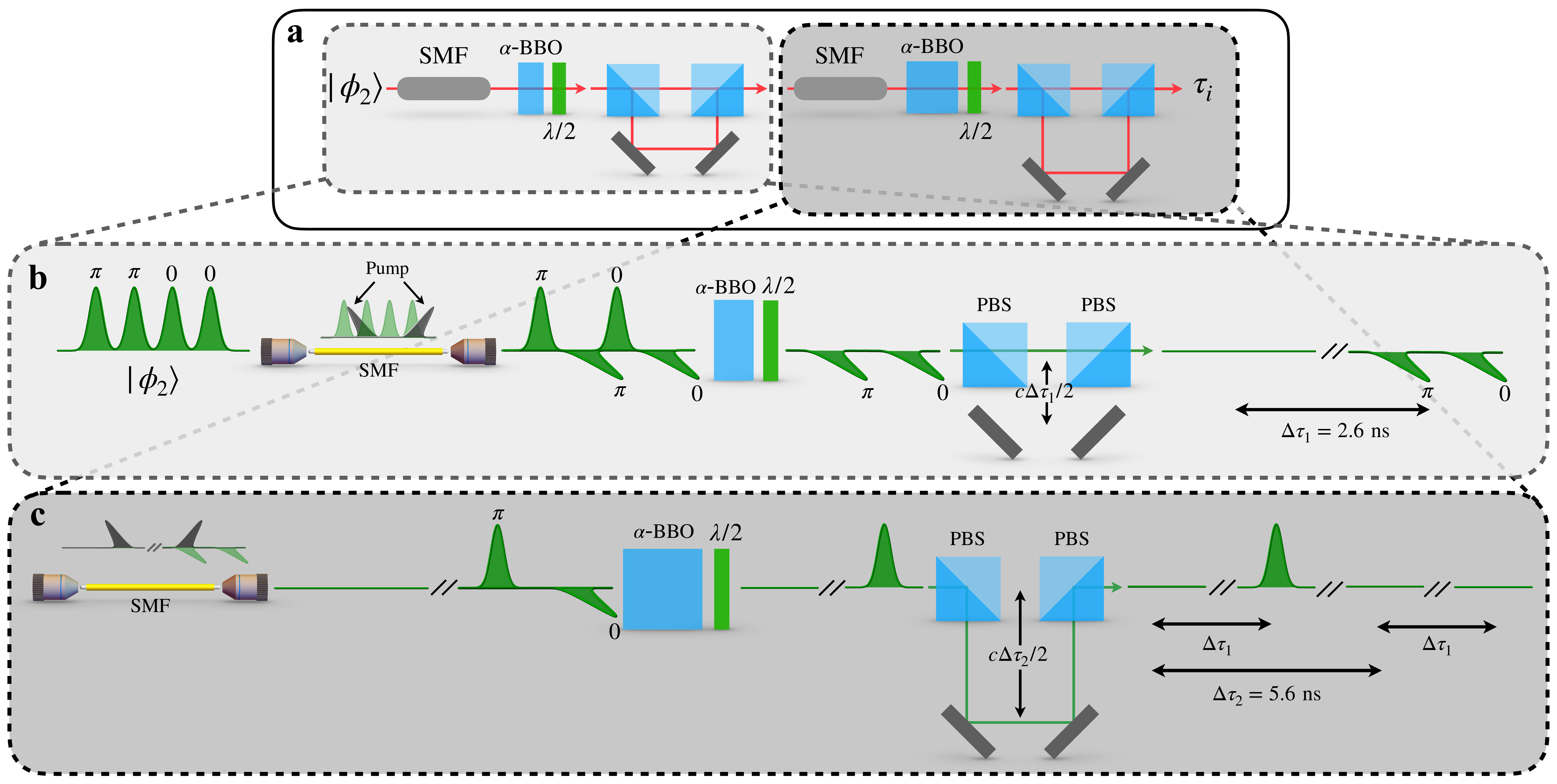}
	\caption{\textbf{Measuring time-bin qudits in the superposition basis with input state $|\phi_2\rangle$}.  \textbf{(a)} Simplified experimental setup for the measurement of superposition time-bin qudits in dimension $d=4$ for input state $|\phi_2\rangle$. \textbf{(b)} The first stage consists of an SMF, a 5-mm $\alpha$-BBO crystal, and a polarization time delay with a path difference of $\sim80$~cm. \textbf{(c)} The second stage consists of an SMF, a 10-mm $\alpha$-BBO crystal, and a polarization time delay with a path difference of $\sim170$~cm.}
	\label{fig:setup3}
\end{figure*}

\begin{figure*}[t!]
	\centering
		\includegraphics[width=0.92\textwidth]{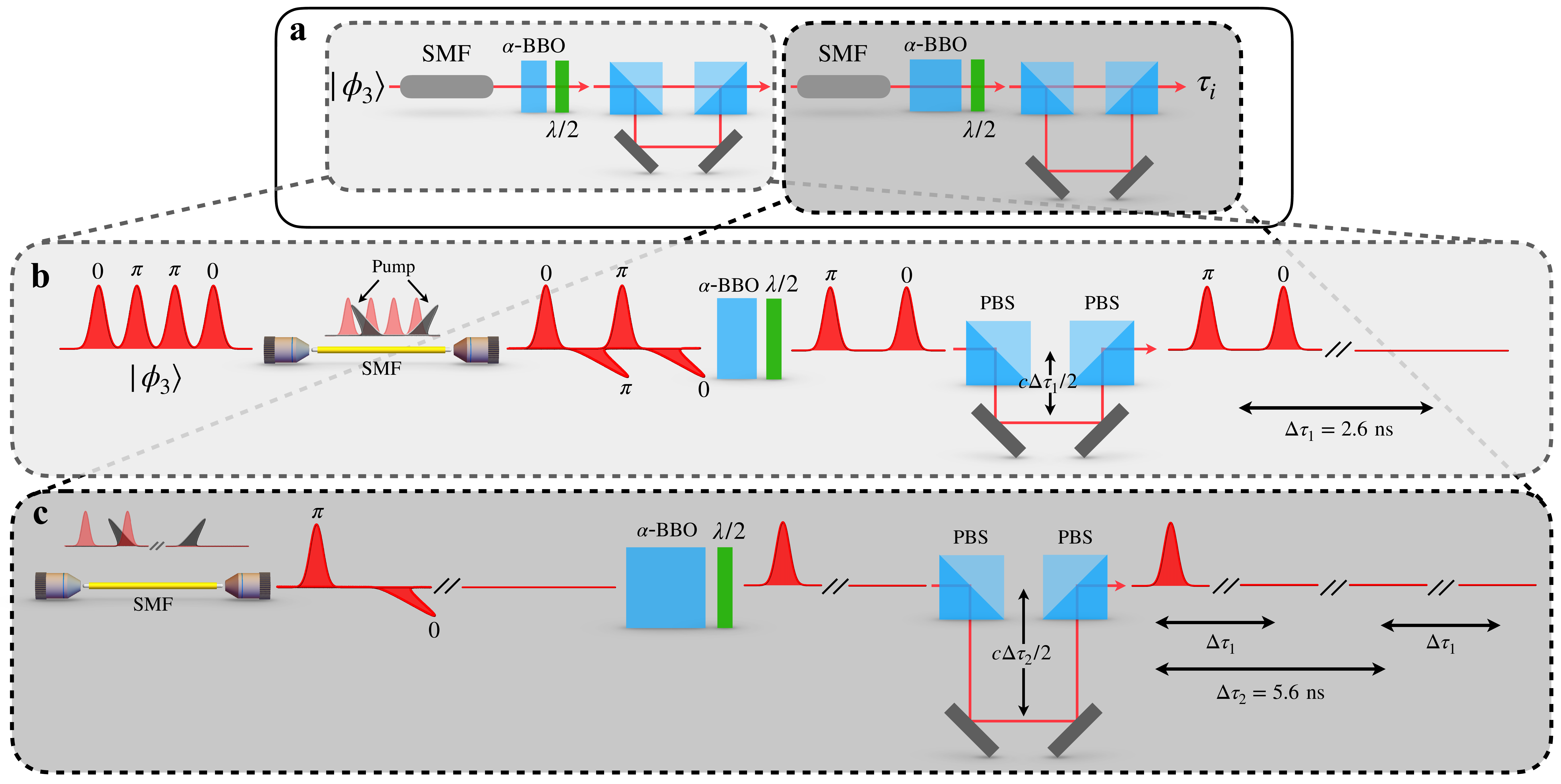}
	\caption{\textbf{Measuring time-bin qudits in the superposition basis with input state $|\phi_3\rangle$}.  \textbf{(a)} Simplified experimental setup for the measurement of superposition time-bin qudits in dimension $d=4$ for input state $|\phi_3\rangle$. \textbf{(b)} The first stage consists of an SMF, a 5-mm $\alpha$-BBO crystal, and a polarization time delay with a path difference of $\sim80$~cm. \textbf{(c)} The second stage consists of an SMF, a 10-mm $\alpha$-BBO crystal, and a polarization time delay with a path difference of $\sim170$~cm.}
	\label{fig:setup4}
\end{figure*}

In Fig.~\ref{fig:setup2}-\ref{fig:setup4}, we show how superposition states $|\phi_1\rangle$, $|\phi_2\rangle$, and $|\phi_3\rangle$ propagate through the measurement apparatus. Note that the apparatus is identical to Fig.~\ref{fig:setup}, but, due to interference, the different input modes are sorted into different temporal bins.

\begin{thebibliography}{37}%
\makeatletter
\providecommand \@ifxundefined [1]{%
 \@ifx{#1\undefined}
}%
\providecommand \@ifnum [1]{%
 \ifnum #1\expandafter \@firstoftwo
 \else \expandafter \@secondoftwo
 \fi
}%
\providecommand \@ifx [1]{%
 \ifx #1\expandafter \@firstoftwo
 \else \expandafter \@secondoftwo
 \fi
}%
\providecommand \natexlab [1]{#1}%
\providecommand \enquote  [1]{``#1''}%
\providecommand \bibnamefont  [1]{#1}%
\providecommand \bibfnamefont [1]{#1}%
\providecommand \citenamefont [1]{#1}%
\providecommand \href@noop [0]{\@secondoftwo}%
\providecommand \href [0]{\begingroup \@sanitize@url \@href}%
\providecommand \@href[1]{\@@startlink{#1}\@@href}%
\providecommand \@@href[1]{\endgroup#1\@@endlink}%
\providecommand \@sanitize@url [0]{\catcode `\\12\catcode `\$12\catcode
  `\&12\catcode `\#12\catcode `\^12\catcode `\_12\catcode `\%12\relax}%
\providecommand \@@startlink[1]{}%
\providecommand \@@endlink[0]{}%
\providecommand \url  [0]{\begingroup\@sanitize@url \@url }%
\providecommand \@url [1]{\endgroup\@href {#1}{\urlprefix }}%
\providecommand \urlprefix  [0]{URL }%
\providecommand \Eprint [0]{\href }%
\providecommand \doibase [0]{http://dx.doi.org/}%
\providecommand \selectlanguage [0]{\@gobble}%
\providecommand \bibinfo  [0]{\@secondoftwo}%
\providecommand \bibfield  [0]{\@secondoftwo}%
\providecommand \translation [1]{[#1]}%
\providecommand \BibitemOpen [0]{}%
\providecommand \bibitemStop [0]{}%
\providecommand \bibitemNoStop [0]{.\EOS\space}%
\providecommand \EOS [0]{\spacefactor3000\relax}%
\providecommand \BibitemShut  [1]{\csname bibitem#1\endcsname}%
\let\auto@bib@innerbib\@empty
\bibitem [{\citenamefont {Dowling}\ and\ \citenamefont
  {Milburn}(2003)}]{dowling2003quantum}%
  \BibitemOpen
  \bibfield  {author} {\bibinfo {author} {\bibfnamefont {J.~P.}\
  \bibnamefont {Dowling}}\ and\ \bibinfo {author} { \bibfnamefont {G.~J.}\
  \bibnamefont {Milburn}},\ }\bibfield  {title} { \textit {\bibinfo {title}
  {Quantum technology: the second quantum revolution},}\ }\href@noop {}
  {\bibfield  {journal} {\bibinfo  {journal} {Phil. Trans. R. Soc. A}\ }\textbf {\bibinfo {volume} {361}},\ \bibinfo {pages} {1655--1674}
  (\bibinfo {year} {2003})}\BibitemShut {NoStop}%
\bibitem [{\citenamefont {Flamini}\ \emph {et~al.}(2018)\citenamefont
  {Flamini}, \citenamefont {Spagnolo},\ and\ \citenamefont
  {Sciarrino}}]{flamini2018photonic}%
  \BibitemOpen
  \bibfield  {author} {\bibinfo {author} {\bibfnamefont {F.}\ \bibnamefont
  {Flamini}}, \bibinfo {author} {\bibfnamefont {N.}\ \bibnamefont
  {Spagnolo}}, and\ \bibinfo {author} {\bibfnamefont {F.}\ \bibnamefont
  {Sciarrino}},\ }\bibfield  {title} {\textit {\bibinfo {title} {Photonic
  quantum information processing: a review},}\ }\href@noop {} {\bibfield
  {journal} {\bibinfo  {journal} {Rep. Prog. Phys.}\ }\textbf
  {\bibinfo {volume} {82}},\ \bibinfo {pages} {016001} (\bibinfo {year}
  {2018})}\BibitemShut {NoStop}%
\bibitem [{\citenamefont {Scarani}\ \emph {et~al.}(2009)\citenamefont
  {Scarani}, \citenamefont {Bechmann-Pasquinucci}, \citenamefont {Cerf},
  \citenamefont {Du\ifmmode~\check{s}\else \v{s}\fi{}ek}, \citenamefont
  {L\"utkenhaus},\ and\ \citenamefont {Peev}}]{scarani:09}%
  \BibitemOpen
  \bibfield  {author} {\bibinfo {author} {\bibfnamefont {V.}\ \bibnamefont
  {Scarani}}, \bibinfo {author} {\bibfnamefont {H.}\ \bibnamefont
  {Bechmann-Pasquinucci}}, \bibinfo {author} {\bibfnamefont {N.~J.}\
  \bibnamefont {Cerf}}, \bibinfo {author} {\bibfnamefont {M.}\
  \bibnamefont {Du\ifmmode~\check{s}\else \v{s}\fi{}ek}}, \bibinfo {author}
  {\bibfnamefont {N.}\ \bibnamefont {L\"utkenhaus}}, \ and\ \bibinfo
  {author} {\bibfnamefont {M.}\ \bibnamefont {Peev}},\ }\bibfield
  {title} {\textit {\bibinfo {title} {The security of practical quantum key
  distribution},}\ }\href {} {\bibfield
  {journal} {\bibinfo  {journal} {Rev. Mod. Phys.}\ }\textbf {\bibinfo {volume}
  {81}},\ \bibinfo {pages} {1301--1350} (\bibinfo {year} {2009})}\BibitemShut
  {NoStop}%
\bibitem [{\citenamefont {Pirandola}\ \emph {et~al.}(2018)\citenamefont
  {Pirandola}, \citenamefont {Bardhan}, \citenamefont {Gehring}, \citenamefont
  {Weedbrook},\ and\ \citenamefont {Lloyd}}]{pirandola2018advances}%
  \BibitemOpen
  \bibfield  {author} {\bibinfo {author} {\bibfnamefont {S.}\ \bibnamefont
  {Pirandola}}, \bibinfo {author} {\bibfnamefont {B.~R.}\ \bibnamefont
  {Bardhan}}, \bibinfo {author} {\bibfnamefont {T.}\ \bibnamefont
  {Gehring}}, \bibinfo {author} {\bibfnamefont {C.}\ \bibnamefont
  {Weedbrook}}, \ and\ \bibinfo {author} {\bibfnamefont {S.}\ \bibnamefont
  {Lloyd}},\ }\bibfield  {title} {\textit {\bibinfo {title} {Advances in
  photonic quantum sensing},}\ }\href@noop {} {\bibfield  {journal} {\bibinfo
  {journal} {Nat. Phot.}\ }\textbf {\bibinfo {volume} {12}},\ \bibinfo
  {pages} {724--733} (\bibinfo {year} {2018})}\BibitemShut {NoStop}%
\bibitem [{\citenamefont {Kwiat}\ \emph {et~al.}(1995)\citenamefont {Kwiat},
  \citenamefont {Mattle}, \citenamefont {Weinfurter}, \citenamefont
  {Zeilinger}, \citenamefont {Sergienko},\ and\ \citenamefont
  {Shih}}]{kwiat1995new}%
  \BibitemOpen
  \bibfield  {author} {\bibinfo {author} {\bibfnamefont {P.~G.}\ \bibnamefont
  {Kwiat}}, \bibinfo {author} {\bibfnamefont {K.}\ \bibnamefont {Mattle}},
  \bibinfo {author} {\bibfnamefont {H.}\ \bibnamefont {Weinfurter}},
  \bibinfo {author} {\bibfnamefont {A.}\ \bibnamefont {Zeilinger}}, \bibinfo
  {author} {\bibfnamefont {A.~V.}\ \bibnamefont {Sergienko}}, \ and\
  \bibinfo {author} {\bibfnamefont {Y.}\ \bibnamefont {Shih}},\ }\bibfield
  {title} {\textit {\bibinfo {title} {New high-intensity source of
  polarization-entangled photon pairs},}\ }\href@noop {} {\bibfield  {journal}
  {\bibinfo  {journal} {Phys. Rev. Lett.}\ }\textbf {\bibinfo {volume}
  {75}},\ \bibinfo {pages} {4337} (\bibinfo {year} {1995})}\BibitemShut
  {NoStop}%
\bibitem [{\citenamefont {Brendel}\ \emph {et~al.}(1999)\citenamefont
  {Brendel}, \citenamefont {Gisin}, \citenamefont {Tittel},\ and\ \citenamefont
  {Zbinden}}]{brendel1999pulsed}%
  \BibitemOpen
  \bibfield  {author} {\bibinfo {author} {\bibfnamefont {J.}\
  \bibnamefont {Brendel}}, \bibinfo {author} {\bibfnamefont {N.}\
  \bibnamefont {Gisin}}, \bibinfo {author} {\bibfnamefont {W.}\
  \bibnamefont {Tittel}}, \ and\ \bibinfo {author} {\bibfnamefont {H.}\
  \bibnamefont {Zbinden}},\ }\bibfield  {title} {\textit {\bibinfo {title}
  {Pulsed energy-time entangled twin-photon source for quantum
  communication},}\ }\href@noop {} {\bibfield  {journal} {\bibinfo  {journal}
  {Phys. Rev. Lett.}\ }\textbf {\bibinfo {volume} {82}},\ \bibinfo
  {pages} {2594} (\bibinfo {year} {1999})}\BibitemShut {NoStop}%
\bibitem [{\citenamefont {Donohue}\ \emph {et~al.}(2013)\citenamefont
  {Donohue}, \citenamefont {Agnew}, \citenamefont {Lavoie},\ and\ \citenamefont
  {Resch}}]{donohue2013coherent}%
  \BibitemOpen
  \bibfield  {author} {\bibinfo {author} {\bibfnamefont {J.~M.}\ \bibnamefont
  {Donohue}}, \bibinfo {author} {\bibfnamefont {M.}\ \bibnamefont {Agnew}},
  \bibinfo {author} {\bibfnamefont {J.}\ \bibnamefont {Lavoie}}, \ and\
  \bibinfo {author} {\bibfnamefont {K.~J.}\ \bibnamefont {Resch}},\
  }\bibfield  {title} {\textit {\bibinfo {title} {Coherent ultrafast
  measurement of time-bin encoded photons},}\ }\href@noop {} {\bibfield
  {journal} {\bibinfo  {journal} {Phys. Rev. Lett.}\ }\textbf {\bibinfo
  {volume} {111}},\ \bibinfo {pages} {153602} (\bibinfo {year}
  {2013})}\BibitemShut {NoStop}%
\bibitem [{\citenamefont {Lukens}\ and\ \citenamefont
  {Lougovski}(2017)}]{lukens2017frequency}%
  \BibitemOpen
  \bibfield  {author} {\bibinfo {author} {\bibfnamefont {J.~M.}\
  \bibnamefont {Lukens}}\ and\ \bibinfo {author} {\bibfnamefont {P.}\
  \bibnamefont {Lougovski}},\ }\bibfield  {title} {\textit {\bibinfo {title}
  {Frequency-encoded photonic qubits for scalable quantum information
  processing},}\ }\href@noop {} {\bibfield  {journal} {\bibinfo  {journal}
  {Optica}\ }\textbf {\bibinfo {volume} {4}},\ \bibinfo {pages} {8--16}
  (\bibinfo {year} {2017})}\BibitemShut {NoStop}%
\bibitem [{\citenamefont {Cari{\~n}e}\ \emph {et~al.}(2020)\citenamefont
  {Cari{\~n}e}, \citenamefont {Ca{\~n}as}, \citenamefont {Skrzypczyk},
  \citenamefont {{\v{S}}upi{\'c}}, \citenamefont {Guerrero}, \citenamefont
  {Garcia}, \citenamefont {Pereira}, \citenamefont {Prosser}, \citenamefont
  {Xavier}, \citenamefont {Delgado} \emph {et~al.}}]{carine2020multi}%
  \BibitemOpen
  \bibfield  {author} {\bibinfo {author} {\bibfnamefont {J.}~\bibnamefont
  {Cari{\~n}e}}, \bibinfo {author} {\bibfnamefont {G.}~\bibnamefont
  {Ca{\~n}as}}, \bibinfo {author} {\bibfnamefont {P.}~\bibnamefont
  {Skrzypczyk}}, \bibinfo {author} {\bibfnamefont {I.}~\bibnamefont
  {{\v{S}}upi{\'c}}}, \bibinfo {author} {\bibfnamefont {N.}~\bibnamefont
  {Guerrero}}, \bibinfo {author} {\bibfnamefont {T.}~\bibnamefont {Garcia}},
  \bibinfo {author} {\bibfnamefont {L.}~\bibnamefont {Pereira}}, \bibinfo
  {author} {\bibfnamefont {M. A. S.}\ \bibnamefont {Prosser}}, \bibinfo {author}
  {\bibfnamefont {G.~B.}\ \bibnamefont {Xavier}}, \bibinfo {author}
  {\bibfnamefont {A.}~\bibnamefont {Delgado}},  \emph {et~al.},\ }\bibfield
  {title} {\textit {\bibinfo {title} {Multi-core fiber integrated multi-port
  beam splitters for quantum information processing},}\ }\href@noop {}
  {\bibfield  {journal} {\bibinfo  {journal} {Optica}\ }\textbf {\bibinfo
  {volume} {7}},\ \bibinfo {pages} {542--550} (\bibinfo {year}
  {2020})}\BibitemShut {NoStop}%
\bibitem [{\citenamefont {Bouchard}\ \emph
  {et~al.}(2018{\natexlab{a}})\citenamefont {Bouchard}, \citenamefont
  {Valencia}, \citenamefont {Brandt}, \citenamefont {Fickler}, \citenamefont
  {Huber},\ and\ \citenamefont {Malik}}]{bouchard2018measuring}%
  \BibitemOpen
  \bibfield  {author} {\bibinfo {author} {\bibfnamefont {F.}\
  \bibnamefont {Bouchard}}, \bibinfo {author} {\bibfnamefont {N.~H.}\
  \bibnamefont {Valencia}}, \bibinfo {author} {\bibfnamefont {F.}\
  \bibnamefont {Brandt}}, \bibinfo {author} {\bibfnamefont {R.}\
  \bibnamefont {Fickler}}, \bibinfo {author} {\bibfnamefont {M.}\
  \bibnamefont {Huber}}, \ and\ \bibinfo {author} {\bibfnamefont {M.}\
  \bibnamefont {Malik}},\ }\bibfield  {title} {\textit {\bibinfo {title}
  {Measuring azimuthal and radial modes of photons},}\ }\href@noop {}
  {\bibfield  {journal} {\bibinfo  {journal} {Opt. Express}\ }\textbf
  {\bibinfo {volume} {26}},\ \bibinfo {pages} {31925--31941} (\bibinfo {year}
  {2018}{\natexlab{a}})}\BibitemShut {NoStop}%
\bibitem [{\citenamefont {Brecht}\ \emph {et~al.}(2015)\citenamefont {Brecht},
  \citenamefont {Reddy}, \citenamefont {Silberhorn},\ and\ \citenamefont
  {Raymer}}]{brecht2015photon}%
  \BibitemOpen
  \bibfield  {author} {\bibinfo {author} {\bibfnamefont {B.}~\bibnamefont
  {Brecht}}, \bibinfo {author} {\bibfnamefont {D.~V.}\ \bibnamefont
  {Reddy}}, \bibinfo {author} {\bibfnamefont {C.}~\bibnamefont {Silberhorn}}, \
  and\ \bibinfo {author} {\bibfnamefont {M. G.}~\bibnamefont {Raymer}},\
  }\bibfield  {title} {\textit {\bibinfo {title} {Photon temporal modes: a
  complete framework for quantum information science},}\ }\href@noop {}
  {\bibfield  {journal} {\bibinfo  {journal} {Phys. Rev. X}\ }\textbf
  {\bibinfo {volume} {5}},\ \bibinfo {pages} {041017} (\bibinfo {year}
  {2015})}\BibitemShut {NoStop}%
\bibitem [{\citenamefont {Braunstein}\ and\ \citenamefont
  {Van~Loock}(2005)}]{braunstein2005quantum}%
  \BibitemOpen
  \bibfield  {author} {\bibinfo {author} {\bibfnamefont {S.~L.}\
  \bibnamefont {Braunstein}}\ and\ \bibinfo {author} {\bibfnamefont {P.}\
  \bibnamefont {Van~Loock}},\ }\bibfield  {title} {\textit {\bibinfo {title}
  {Quantum information with continuous variables},}\ }\href@noop {} {\bibfield
  {journal} {\bibinfo  {journal} {Rev. Mod. Phys.}\ }\textbf
  {\bibinfo {volume} {77}},\ \bibinfo {pages} {513} (\bibinfo {year}
  {2005})}\BibitemShut {NoStop}%
\bibitem [{\citenamefont {Kues}\ \emph {et~al.}(2017)\citenamefont {Kues},
  \citenamefont {Reimer}, \citenamefont {Roztocki}, \citenamefont {Cort{\'e}s},
  \citenamefont {Sciara}, \citenamefont {Wetzel}, \citenamefont {Zhang},
  \citenamefont {Cino}, \citenamefont {Chu}, \citenamefont {Little} \emph
  {et~al.}}]{kues2017chip}%
  \BibitemOpen
  \bibfield  {author} {\bibinfo {author} {\bibfnamefont {M.}\ \bibnamefont
  {Kues}}, \bibinfo {author} {\bibfnamefont {C.}\ \bibnamefont
  {Reimer}}, \bibinfo {author} {\bibfnamefont {P.}\ \bibnamefont
  {Roztocki}}, \bibinfo {author} {\bibfnamefont {L.~R.}\ \bibnamefont
  {Cort{\'e}s}}, \bibinfo {author} {\bibfnamefont {S.}\ \bibnamefont
  {Sciara}}, \bibinfo {author} {\bibfnamefont {B.}\ \bibnamefont
  {Wetzel}}, \bibinfo {author} {\bibfnamefont {Y.}\ \bibnamefont {Zhang}},
  \bibinfo {author} {\bibfnamefont {A.}\ \bibnamefont {Cino}}, \bibinfo
  {author} {\bibfnamefont {S.~T.}\ \bibnamefont {Chu}}, \bibinfo {author}
  {\bibfnamefont {B.~E.}\ \bibnamefont {Little}},  \emph {et~al.},\
  }\bibfield  {title} {\textit {\bibinfo {title} {On-chip generation of
  high-dimensional entangled quantum states and their coherent control},}\
  }\href@noop {} {\bibfield  {journal} {\bibinfo  {journal} {Nature}\ }\textbf
  {\bibinfo {volume} {546}},\ \bibinfo {pages} {622--626} (\bibinfo {year}
  {2017})}\BibitemShut {NoStop}%
\bibitem [{\citenamefont {Wang}\ \emph {et~al.}(2018)\citenamefont {Wang},
  \citenamefont {Paesani}, \citenamefont {Ding}, \citenamefont {Santagati},
  \citenamefont {Skrzypczyk}, \citenamefont {Salavrakos}, \citenamefont {Tura},
  \citenamefont {Augusiak}, \citenamefont {Man{\v{c}}inska}, \citenamefont
  {Bacco} \emph {et~al.}}]{wang2018multidimensional}%
  \BibitemOpen
  \bibfield  {author} {\bibinfo {author} {\bibfnamefont {J.}\ \bibnamefont
  {Wang}}, \bibinfo {author} {\bibfnamefont {S.}\ \bibnamefont {Paesani}},
  \bibinfo {author} {\bibfnamefont {Y.}\ \bibnamefont {Ding}}, \bibinfo
  {author} {\bibfnamefont {R.}\ \bibnamefont {Santagati}}, \bibinfo
  {author} {\bibfnamefont {P.}\ \bibnamefont {Skrzypczyk}}, \bibinfo {author}
  {\bibfnamefont {A.}\ \bibnamefont {Salavrakos}}, \bibinfo {author}
  {\bibfnamefont {J.}\ \bibnamefont {Tura}}, \bibinfo {author}
  {\bibfnamefont {R.}\ \bibnamefont {Augusiak}}, \bibinfo {author}
  {\bibfnamefont {L.}\ \bibnamefont {Man{\v{c}}inska}}, \bibinfo {author}
  {\bibfnamefont {D.}\ \bibnamefont {Bacco}},  \emph {et~al.},\ }\bibfield
  {title} {\textit {\bibinfo {title} {Multidimensional quantum entanglement
  with large-scale integrated optics},}\ }\href@noop {} {\bibfield  {journal}
  {\bibinfo  {journal} {Science}\ }\textbf {\bibinfo {volume} {360}},\ \bibinfo
  {pages} {285--291} (\bibinfo {year} {2018})}\BibitemShut {NoStop}%
\bibitem [{\citenamefont {Malik}\ \emph {et~al.}(2016)\citenamefont {Malik},
  \citenamefont {Erhard}, \citenamefont {Huber}, \citenamefont {Krenn},
  \citenamefont {Fickler},\ and\ \citenamefont {Zeilinger}}]{malik2016multi}%
  \BibitemOpen
  \bibfield  {author} {\bibinfo {author} {\bibfnamefont {M.}\ \bibnamefont
  {Malik}}, \bibinfo {author} {\bibfnamefont {M.}\ \bibnamefont {Erhard}},
  \bibinfo {author} {\bibfnamefont {M.}\ \bibnamefont {Huber}}, \bibinfo
  {author} {\bibfnamefont {M.}\ \bibnamefont {Krenn}}, \bibinfo {author}
  {\bibfnamefont {R.}\ \bibnamefont {Fickler}}, \ and\ \bibinfo {author}
  {\bibfnamefont {A.}\ \bibnamefont {Zeilinger}},\ }\bibfield  {title}
  {\textit {\bibinfo {title} {Multi-photon entanglement in high dimensions},}\
  }\href@noop {} {\bibfield  {journal} {\bibinfo  {journal} {Nat. Phot.}\
  }\textbf {\bibinfo {volume} {10}},\ \bibinfo {pages} {248--252} (\bibinfo
  {year} {2016})}\BibitemShut {NoStop}%
\bibitem [{\citenamefont {Erhard}\ \emph {et~al.}(2020)\citenamefont {Erhard},
  \citenamefont {Krenn},\ and\ \citenamefont {Zeilinger}}]{erhard2020advances}%
  \BibitemOpen
  \bibfield  {author} {\bibinfo {author} {\bibfnamefont {M.}\ \bibnamefont
  {Erhard}}, \bibinfo {author} {\bibfnamefont {M.}\ \bibnamefont {Krenn}}, \
  and\ \bibinfo {author} {\bibfnamefont {A.}\ \bibnamefont {Zeilinger}},\
  }\bibfield  {title} {\textit {\bibinfo {title} {Advances in high-dimensional
  quantum entanglement},}\ }\href@noop {} {\bibfield  {journal} {\bibinfo
  {journal} {Nat. Rev. Phys.}\ }\textbf {\bibinfo {volume} {2}},\
  \bibinfo {pages} {365--381} (\bibinfo {year} {2020})}\BibitemShut {NoStop}%
\bibitem [{\citenamefont {Mower}\ \emph {et~al.}(2013)\citenamefont {Mower},
  \citenamefont {Zhang}, \citenamefont {Desjardins}, \citenamefont {Lee},
  \citenamefont {Shapiro},\ and\ \citenamefont {Englund}}]{mower2013high}%
  \BibitemOpen
  \bibfield  {author} {\bibinfo {author} {\bibfnamefont {J.}\ \bibnamefont
  {Mower}}, \bibinfo {author} {\bibfnamefont {Z.}\ \bibnamefont {Zhang}},
  \bibinfo {author} {\bibfnamefont {P.}\ \bibnamefont {Desjardins}},
  \bibinfo {author} {\bibfnamefont {C.}\ \bibnamefont {Lee}}, \bibinfo
  {author} {\bibfnamefont {J.~H.}\ \bibnamefont {Shapiro}}, \ and\ \bibinfo
  {author} {\bibfnamefont {D.}\ \bibnamefont {Englund}},\ }\bibfield  {title}
  {\textit {\bibinfo {title} {High-dimensional quantum key distribution using
  dispersive optics},}\ }\href@noop {} {\bibfield  {journal} {\bibinfo
  {journal} {Phys. Rev. A}\ }\textbf {\bibinfo {volume} {87}},\ \bibinfo
  {pages} {062322} (\bibinfo {year} {2013})}\BibitemShut {NoStop}%
\bibitem [{\citenamefont {Ecker}\ \emph {et~al.}(2019)\citenamefont {Ecker},
  \citenamefont {Bouchard}, \citenamefont {Bulla}, \citenamefont {Brandt},
  \citenamefont {Kohout}, \citenamefont {Steinlechner}, \citenamefont
  {Fickler}, \citenamefont {Malik}, \citenamefont {Guryanova}, \citenamefont
  {Ursin},\ and\ \citenamefont {Huber}}]{ecker2019overcoming}%
  \BibitemOpen
  \bibfield  {author} {\bibinfo {author} {\bibfnamefont {S.}\
  \bibnamefont {Ecker}}, \bibinfo {author} {\bibfnamefont {F.}\
  \bibnamefont {Bouchard}}, \bibinfo {author} {\bibfnamefont {L.}\
  \bibnamefont {Bulla}}, \bibinfo {author} {\bibfnamefont {F.}\
  \bibnamefont {Brandt}}, \bibinfo {author} {\bibfnamefont {O.}\
  \bibnamefont {Kohout}}, \bibinfo {author} {\bibfnamefont {F.}\
  \bibnamefont {Steinlechner}}, \bibinfo {author} {\bibfnamefont {R.}\
  \bibnamefont {Fickler}}, \bibinfo {author} {\bibfnamefont {M.}\
  \bibnamefont {Malik}}, \bibinfo {author} {\bibfnamefont {Y.}\
  \bibnamefont {Guryanova}}, \bibinfo {author} {\bibfnamefont {R.}\
  \bibnamefont {Ursin}}, \ and\ \bibinfo {author} {\bibfnamefont {M.}\
  \bibnamefont {Huber}},\ }\bibfield  {title} {\textit {\bibinfo {title}
  {Overcoming noise in entanglement distribution},}\ }\href {\doibase
  10.1103/PhysRevX.9.041042} {\bibfield  {journal} {\bibinfo  {journal} {Phys.
  Rev. X}\ }\textbf {\bibinfo {volume} {9}},\ \bibinfo {pages} {041042}
  (\bibinfo {year} {2019})}\BibitemShut {NoStop}%
\bibitem [{\citenamefont {Clements}\ \emph {et~al.}(2016)\citenamefont
  {Clements}, \citenamefont {Humphreys}, \citenamefont {Metcalf}, \citenamefont
  {Kolthammer},\ and\ \citenamefont {Walmsley}}]{clements2016optimal}%
  \BibitemOpen
  \bibfield  {author} {\bibinfo {author} {\bibfnamefont {W.~R.}\
  \bibnamefont {Clements}}, \bibinfo {author} {\bibfnamefont {P.~C.}\
  \bibnamefont {Humphreys}}, \bibinfo {author} {\bibfnamefont {B.~J.}\
  \bibnamefont {Metcalf}}, \bibinfo {author} {\bibfnamefont {W.~S.}\
  \bibnamefont {Kolthammer}}, \ and\ \bibinfo {author} {\bibfnamefont {I.~A.}\
  \bibnamefont {Walmsley}},\ }\bibfield  {title} {\textit {\bibinfo {title}
  {Optimal design for universal multiport interferometers},}\ }\href@noop {}
  {\bibfield  {journal} {\bibinfo  {journal} {Optica}\ }\textbf {\bibinfo
  {volume} {3}},\ \bibinfo {pages} {1460--1465} (\bibinfo {year}
  {2016})}\BibitemShut {NoStop}%
\bibitem [{\citenamefont {Marcikic}\ \emph {et~al.}(2002)\citenamefont
  {Marcikic}, \citenamefont {de~Riedmatten}, \citenamefont {Tittel},
  \citenamefont {Scarani}, \citenamefont {Zbinden},\ and\ \citenamefont
  {Gisin}}]{marcikic2002time}%
  \BibitemOpen
  \bibfield  {author} {\bibinfo {author} {\bibfnamefont {I.}\ \bibnamefont
  {Marcikic}}, \bibinfo {author} {\bibfnamefont {H.}\ \bibnamefont
  {de~Riedmatten}}, \bibinfo {author} {\bibfnamefont {W.}\ \bibnamefont
  {Tittel}}, \bibinfo {author} {\bibfnamefont {V.}\ \bibnamefont
  {Scarani}}, \bibinfo {author} {\bibfnamefont {H.}\ \bibnamefont {Zbinden}},
  \ and\ \bibinfo {author} {\bibfnamefont {N.}\ \bibnamefont {Gisin}},\
  }\bibfield  {title} {\textit {\bibinfo {title} {Time-bin entangled qubits
  for quantum communication created by femtosecond pulses},}\ }\href@noop {}
  {\bibfield  {journal} {\bibinfo  {journal} {Phys. Rev. A}\ }\textbf
  {\bibinfo {volume} {66}},\ \bibinfo {pages} {062308} (\bibinfo {year}
  {2002})}\BibitemShut {NoStop}%
\bibitem [{\citenamefont {Brougham}\ \emph {et~al.}(2013)\citenamefont
  {Brougham}, \citenamefont {Barnett}, \citenamefont {McCusker}, \citenamefont
  {Kwiat},\ and\ \citenamefont {Gauthier}}]{brougham2013security}%
  \BibitemOpen
  \bibfield  {author} {\bibinfo {author} {\bibfnamefont {T.}\ \bibnamefont
  {Brougham}}, \bibinfo {author} {\bibfnamefont {S.~M.}\ \bibnamefont
  {Barnett}}, \bibinfo {author} {\bibfnamefont {K.~T.}\ \bibnamefont
  {McCusker}}, \bibinfo {author} {\bibfnamefont {P.~G.}\ \bibnamefont
  {Kwiat}}, \ and\ \bibinfo {author} {\bibfnamefont {D.~J.}\ \bibnamefont
  {Gauthier}},\ }\bibfield  {title} {\textit {\bibinfo {title} {Security of
  high-dimensional quantum key distribution protocols using franson
  interferometers},}\ }\href@noop {} {\bibfield  {journal} {\bibinfo  {journal}
  {J. Phys. B}\ }\textbf
  {\bibinfo {volume} {46}},\ \bibinfo {pages} {104010} (\bibinfo {year}
  {2013})}\BibitemShut {NoStop}%
\bibitem [{\citenamefont {Islam}\ \emph {et~al.}(2017)\citenamefont {Islam},
  \citenamefont {Lim}, \citenamefont {Cahall}, \citenamefont {Kim},\ and\
  \citenamefont {Gauthier}}]{islam2017provably}%
  \BibitemOpen
  \bibfield  {author} {\bibinfo {author} {\bibfnamefont {N.~T.}\ \bibnamefont
  {Islam}}, \bibinfo {author} {\bibfnamefont {C. C.~W.}\ \bibnamefont
  {Lim}}, \bibinfo {author} {\bibfnamefont {C.}\ \bibnamefont {Cahall}},
  \bibinfo {author} {\bibfnamefont {J.}\ \bibnamefont {Kim}}, \ and\
  \bibinfo {author} {\bibfnamefont {D.~J.}\ \bibnamefont {Gauthier}},\
  }\bibfield  {title} {\textit {\bibinfo {title} {Provably secure and
  high-rate quantum key distribution with time-bin qudits},}\ }\href@noop {}
  {\bibfield  {journal} {\bibinfo  {journal} {Sci. Adv.}\ }\textbf
  {\bibinfo {volume} {3}},\ \bibinfo {pages} {e1701491} (\bibinfo {year}
  {2017})}\BibitemShut {NoStop}%
\bibitem [{\citenamefont {Ikuta}\ and\ \citenamefont
  {Takesue}(2017)}]{ikuta2017implementation}%
  \BibitemOpen
  \bibfield  {author} {\bibinfo {author} {\bibfnamefont {T.}\ \bibnamefont
  {Ikuta}}\ and\ \bibinfo {author} {\bibfnamefont {H.}\ \bibnamefont
  {Takesue}},\ }\bibfield  {title} {\textit {\bibinfo {title} {Implementation
  of quantum state tomography for time-bin qudits},}\ }\href@noop {} {\bibfield
   {journal} {\bibinfo  {journal} {New J. Phys.}\ }\textbf {\bibinfo
  {volume} {19}},\ \bibinfo {pages} {013039} (\bibinfo {year}
  {2017})}\BibitemShut {NoStop}%
\bibitem [{\citenamefont {Nowierski}\ \emph {et~al.}(2016)\citenamefont
  {Nowierski}, \citenamefont {Oza}, \citenamefont {Kumar},\ and\ \citenamefont
  {Kanter}}]{nowierski2016tomographic}%
  \BibitemOpen
  \bibfield  {author} {\bibinfo {author} {\bibfnamefont {S.~J.}\
  \bibnamefont {Nowierski}}, \bibinfo {author} {\bibfnamefont {N.~N.}\
  \bibnamefont {Oza}}, \bibinfo {author} {\bibfnamefont {P.}\ \bibnamefont
  {Kumar}}, \ and\ \bibinfo {author} {\bibfnamefont {G.~S.}\ \bibnamefont
  {Kanter}},\ }\bibfield  {title} {\textit {\bibinfo {title} {Tomographic
  reconstruction of time-bin-entangled qudits},}\ }\href@noop {} {\bibfield
  {journal} {\bibinfo  {journal} {Phys. Rev. A}\ }\textbf {\bibinfo
  {volume} {94}},\ \bibinfo {pages} {042328} (\bibinfo {year}
  {2016})}\BibitemShut {NoStop}%
\bibitem [{\citenamefont {Kupchak}\ \emph {et~al.}(2017)\citenamefont
  {Kupchak}, \citenamefont {Bustard}, \citenamefont {Heshami}, \citenamefont
  {Erskine}, \citenamefont {Spanner}, \citenamefont {England},\ and\
  \citenamefont {Sussman}}]{kupchak2017time}%
  \BibitemOpen
  \bibfield  {author} {\bibinfo {author} {\bibfnamefont {C.}\ \bibnamefont
  {Kupchak}}, \bibinfo {author} {\bibfnamefont {P.~J.}\ \bibnamefont
  {Bustard}}, \bibinfo {author} {\bibfnamefont {K.}\ \bibnamefont
  {Heshami}}, \bibinfo {author} {\bibfnamefont {J.}\ \bibnamefont
  {Erskine}}, \bibinfo {author} {\bibfnamefont {M.}\ \bibnamefont
  {Spanner}}, \bibinfo {author} {\bibfnamefont {D.~G.}\ \bibnamefont
  {England}}, \ and\ \bibinfo {author} {\bibfnamefont {B.~J.}\
  \bibnamefont {Sussman}},\ }\bibfield  {title} {\textit {\bibinfo {title}
  {Time-bin-to-polarization conversion of ultrafast photonic qubits},}\
  }\href@noop {} {\bibfield  {journal} {\bibinfo  {journal} {Phys. Rev. A}\ }\textbf {\bibinfo {volume} {96}},\ \bibinfo {pages} {053812} (\bibinfo
  {year} {2017})}\BibitemShut {NoStop}%
\bibitem [{\citenamefont {Kupchak}\ \emph {et~al.}(2019)\citenamefont
  {Kupchak}, \citenamefont {Erskine}, \citenamefont {England},\ and\
  \citenamefont {Sussman}}]{kupchak2019terahertz}%
  \BibitemOpen
  \bibfield  {author} {\bibinfo {author} {\bibfnamefont {C.}\ \bibnamefont
  {Kupchak}}, \bibinfo {author} {\bibfnamefont {J.}\ \bibnamefont
  {Erskine}}, \bibinfo {author} {\bibfnamefont {D.}\ \bibnamefont
  {England}}, \ and\ \bibinfo {author} {\bibfnamefont {B.}\ \bibnamefont
  {Sussman}},\ }\bibfield  {title} {\textit {\bibinfo {title}
  {Terahertz-bandwidth switching of heralded single photons},}\ }\href@noop {}
  {\bibfield  {journal} {\bibinfo  {journal} {Opt. Lett.}\ }\textbf
  {\bibinfo {volume} {44}},\ \bibinfo {pages} {1427--1430} (\bibinfo {year}
  {2019})}\BibitemShut {NoStop}%
\bibitem [{\citenamefont {Bouchard}\ \emph {et~al.}(2022)\citenamefont
  {Bouchard}, \citenamefont {England}, \citenamefont {Bustard}, \citenamefont
  {Heshami},\ and\ \citenamefont {Sussman}}]{bouchard2022quantum}%
  \BibitemOpen
  \bibfield  {author} {\bibinfo {author} {\bibfnamefont {F.}\
  \bibnamefont {Bouchard}}, \bibinfo {author} {\bibfnamefont {D.}\
  \bibnamefont {England}}, \bibinfo {author} {\bibfnamefont {P.~J.}\
  \bibnamefont {Bustard}}, \bibinfo {author} {\bibfnamefont {K.}\
  \bibnamefont {Heshami}}, \ and\ \bibinfo {author} {\bibfnamefont {B.}\
  \bibnamefont {Sussman}},\ }\bibfield  {title} {\textit {\bibinfo {title}
  {Quantum communication with ultrafast time-bin qubits},}\ }\href@noop {}
  {\bibfield  {journal} {\bibinfo  {journal} {PRX Quantum}\ }\textbf {\bibinfo
  {volume} {3}},\ \bibinfo {pages} {010332} (\bibinfo {year}
  {2022})}\BibitemShut {NoStop}%
\bibitem [{\citenamefont {Lukens}\ \emph {et~al.}(2018)\citenamefont {Lukens},
  \citenamefont {Islam}, \citenamefont {Lim},\ and\ \citenamefont
  {Gauthier}}]{lukens2018reconfigurable}%
  \BibitemOpen
  \bibfield  {author} {\bibinfo {author} {\bibfnamefont {J.~M.}\
  \bibnamefont {Lukens}}, \bibinfo {author} {\bibfnamefont {N.~T.}\
  \bibnamefont {Islam}}, \bibinfo {author} {\bibfnamefont {C. C.~W.}\
  \bibnamefont {Lim}}, \ and\ \bibinfo {author} {\bibfnamefont {D.~J.}\
  \bibnamefont {Gauthier}},\ }\bibfield  {title} {\textit {\bibinfo {title}
  {Reconfigurable generation and measurement of mutually unbiased bases for
  time-bin qudits},}\ }\href@noop {} {\bibfield  {journal} {\bibinfo  {journal}
  {Appl. Phys. Lett.}\ }\textbf {\bibinfo {volume} {112}},\ \bibinfo
  {pages} {111102} (\bibinfo {year} {2018})}\BibitemShut {NoStop}%
\bibitem [{\citenamefont {Ashby}\ \emph {et~al.}(2020)\citenamefont {Ashby},
  \citenamefont {Thiel}, \citenamefont {Allgaier}, \citenamefont
  {d’Ornellas}, \citenamefont {Davis},\ and\ \citenamefont
  {Smith}}]{ashby2020temporal}%
  \BibitemOpen
  \bibfield  {author} {\bibinfo {author} {\bibfnamefont {J.}\ \bibnamefont
  {Ashby}}, \bibinfo {author} {\bibfnamefont {V.}\ \bibnamefont
  {Thiel}}, \bibinfo {author} {\bibfnamefont {M.}\ \bibnamefont
  {Allgaier}}, \bibinfo {author} {\bibfnamefont {P.}\ \bibnamefont
  {dOrnellas}}, \bibinfo {author} {\bibfnamefont {A.~O. C.}\ \bibnamefont
  {Davis}}, \ and\ \bibinfo {author} {\bibfnamefont {B.~J.}\ \bibnamefont
  {Smith}},\ }\bibfield  {title} {\textit {\bibinfo {title} {Temporal mode
  transformations by sequential time and frequency phase modulation for
  applications in quantum information science},}\ }\href@noop {} {\bibfield
  {journal} {\bibinfo  {journal} {Opt. Exp.}\ }\textbf {\bibinfo {volume}
  {28}},\ \bibinfo {pages} {38376--38389} (\bibinfo {year} {2020})}\BibitemShut
  {NoStop}%
\bibitem [{\citenamefont {Agrawal}(2000)}]{agrawal2000nonlinear}%
  \BibitemOpen
  \bibfield  {author} {\bibinfo {author} {\bibfnamefont {G.~P.}\
  \bibnamefont {Agrawal}},\ }\bibfield  {title} {\textit {\bibinfo {title}
  {Nonlinear fiber optics},}\ }in\ \href@noop {} {\emph {\bibinfo {booktitle}
  {Nonlinear Science at the Dawn of the 21st Century}}}\ (\bibinfo  {publisher}
  {Springer},\ \bibinfo {year} {2000})\ pp.\ \bibinfo {pages}
  {195--211}\BibitemShut {NoStop}%
 \bibitem [{\citenamefont {England}\ and\ \citenamefont
  {Bouchard}(2021)}]{england2021perspectives}%
  \BibitemOpen
  \bibfield  {author} {\bibinfo {author} {\bibfnamefont {D.}\ \bibnamefont
  {England}}, \bibinfo {author} {\bibfnamefont {F.}\ \bibnamefont
  {Bouchard}}, \bibinfo {author} {\bibfnamefont {K.}\ \bibnamefont
  {Fenwick}}, \bibinfo {author} {\bibfnamefont {K.}\ \bibnamefont
  {Bonsma-Fisher}}, \bibinfo {author} {\bibfnamefont {Y.}\ \bibnamefont
  {Zhang}}, \bibinfo {author} {\bibfnamefont {P.}\ \bibnamefont
  {Bustard}}\ and\ \bibinfo {author} {\bibfnamefont {B.}\ \bibnamefont
  {Sussman}},\ }\bibfield  {title} {\textit {\bibinfo {title} {Perspectives on all-optical Kerr switching for quantum optical applications},}\ }\href@noop {}
  {\bibfield  {journal} {\bibinfo  {journal} {Appl. Phys. Lett.}\ }\textbf
  {\bibinfo {volume} {119}},\ \bibinfo {pages} {160501} (\bibinfo {year}
  {2021})}\BibitemShut {NoStop}%
\bibitem [{\citenamefont {Hwang}(2003)}]{hwang2003quantum}%
  \BibitemOpen
  \bibfield  {author} {\bibinfo {author} {\bibfnamefont {W.-Y.}\
  \bibnamefont {Hwang}},\ }\bibfield  {title} {\textit {\bibinfo {title}
  {Quantum key distribution with high loss: toward global secure
  communication},}\ }\href@noop {} {\bibfield  {journal} {\bibinfo  {journal}
  {Phys. Rev. Lett.}\ }\textbf {\bibinfo {volume} {91}},\ \bibinfo
  {pages} {057901} (\bibinfo {year} {2003})}\BibitemShut {NoStop}%
\bibitem [{\citenamefont {Sheridan}\ and\ \citenamefont
  {Scarani}(2010)}]{sheridan2010security}%
  \BibitemOpen
  \bibfield  {author} {\bibinfo {author} {\bibfnamefont {L.}\ \bibnamefont
  {Sheridan}}\ and\ \bibinfo {author} {\bibfnamefont {V.}\ \bibnamefont
  {Scarani}},\ }\bibfield  {title} {\textit {\bibinfo {title} {Security proof
  for quantum key distribution using qudit systems},}\ }\href@noop {}
  {\bibfield  {journal} {\bibinfo  {journal} {Phys. Rev. A}\ }\textbf
  {\bibinfo {volume} {82}},\ \bibinfo {pages} {030301} (\bibinfo {year}
  {2010})}\BibitemShut {NoStop}%
  \bibitem [{\citenamefont {Ma}\ and\ \citenamefont
  {Qi}(2010)}]{ma2005practical}%
  \BibitemOpen
  \bibfield  {author} {\bibinfo {author} {\bibfnamefont {X.}\ \bibnamefont
  {Ma}}, \bibinfo {author} {\bibfnamefont {B.}\ \bibnamefont
  {Qi}}, \bibinfo {author} {\bibfnamefont {Y.}\ \bibnamefont
  {Zhao}}\ and\ \bibinfo {author} {\bibfnamefont {H.-K.}\ \bibnamefont
  {Lo}},\ }\bibfield  {title} {\textit {\bibinfo {title} {Practical decoy state for quantum key distribution},}\ }\href@noop {}
  {\bibfield  {journal} {\bibinfo  {journal} {Phys. Rev. A}\ }\textbf
  {\bibinfo {volume} {72}},\ \bibinfo {pages} {012326} (\bibinfo {year}
  {2005})}\BibitemShut {NoStop}%
\bibitem [{\citenamefont {Sit}\ \emph {et~al.}(2017)\citenamefont {Sit},
  \citenamefont {Bouchard}, \citenamefont {Fickler}, \citenamefont
  {Gagnon-Bischoff}, \citenamefont {Larocque}, \citenamefont {Heshami},
  \citenamefont {Elser}, \citenamefont {Peuntinger}, \citenamefont
  {G{\"u}nthner}, \citenamefont {Heim} \emph {et~al.}}]{sit2017high}%
  \BibitemOpen
  \bibfield  {author} {\bibinfo {author} {\bibfnamefont {A.}\ \bibnamefont
  {Sit}}, \bibinfo {author} {\bibfnamefont {F.}\ \bibnamefont
  {Bouchard}}, \bibinfo {author} {\bibfnamefont {R.}\ \bibnamefont
  {Fickler}}, \bibinfo {author} {\bibfnamefont {J.}\ \bibnamefont
  {Gagnon-Bischoff}}, \bibinfo {author} {\bibfnamefont {H.}\ \bibnamefont
  {Larocque}}, \bibinfo {author} {\bibfnamefont {K.}\ \bibnamefont
  {Heshami}}, \bibinfo {author} {\bibfnamefont {D.}\ \bibnamefont
  {Elser}}, \bibinfo {author} {\bibfnamefont {C.}\ \bibnamefont
  {Peuntinger}}, \bibinfo {author} {\bibfnamefont {K.}\ \bibnamefont
  {G{\"u}nthner}}, \bibinfo {author} {\bibfnamefont {B.}\ \bibnamefont
  {Heim}},  \emph {et~al.},\ }\bibfield  {title} {\textit {\bibinfo {title}
  {High-dimensional intracity quantum cryptography with structured photons},}\
  }\href@noop {} {\bibfield  {journal} {\bibinfo  {journal} {Optica}\ }\textbf
  {\bibinfo {volume} {4}},\ \bibinfo {pages} {1006--1010} (\bibinfo {year}
  {2017})}\BibitemShut {NoStop}%
\bibitem [{\citenamefont {Ding}\ \emph {et~al.}(2017)\citenamefont {Ding},
  \citenamefont {Bacco}, \citenamefont {Dalgaard}, \citenamefont {Cai},
  \citenamefont {Zhou}, \citenamefont {Rottwitt},\ and\ \citenamefont
  {Oxenl{\o}we}}]{ding2017high}%
  \BibitemOpen
  \bibfield  {author} {\bibinfo {author} {\bibfnamefont {Y.}\ \bibnamefont
  {Ding}}, \bibinfo {author} {\bibfnamefont {D.}\ \bibnamefont {Bacco}},
  \bibinfo {author} {\bibfnamefont {K.}\ \bibnamefont {Dalgaard}}, \bibinfo
  {author} {\bibfnamefont {X.}\ \bibnamefont {Cai}}, \bibinfo {author}
  {\bibfnamefont {X.}\ \bibnamefont {Zhou}}, \bibinfo {author}
  {\bibfnamefont {K.}\ \bibnamefont {Rottwitt}}, \ and\ \bibinfo {author}
  {\bibfnamefont {L.~K.}\ \bibnamefont {Oxenl{\o}we}},\ }\bibfield
  {title} {\textit {\bibinfo {title} {High-dimensional quantum key
  distribution based on multicore fiber using silicon photonic integrated
  circuits},}\ }\href@noop {} {\bibfield  {journal} {\bibinfo  {journal} {npj
  Quantum Inf.}\ }\textbf {\bibinfo {volume} {3}},\ \bibinfo {pages}
  {1--7} (\bibinfo {year} {2017})}\BibitemShut {NoStop}%
\bibitem [{\citenamefont {Bouchard}\ \emph
  {et~al.}(2018{\natexlab{b}})\citenamefont {Bouchard}, \citenamefont
  {Heshami}, \citenamefont {England}, \citenamefont {Fickler}, \citenamefont
  {Boyd}, \citenamefont {Englert}, \citenamefont {S{\'{a}}nchez-Soto},\ and\
  \citenamefont {Karimi}}]{bouchard2018experimental}%
  \BibitemOpen
  \bibfield  {author} {\bibinfo {author} {\bibfnamefont {F.}\
  \bibnamefont {Bouchard}}, \bibinfo {author} {\bibfnamefont {K.}\
  \bibnamefont {Heshami}}, \bibinfo {author} {\bibfnamefont {D.}\
  \bibnamefont {England}}, \bibinfo {author} {\bibfnamefont {R.}\
  \bibnamefont {Fickler}}, \bibinfo {author} {\bibfnamefont {R.~W.}\
  \bibnamefont {Boyd}}, \bibinfo {author} {\bibfnamefont {B.-G.}\
  \bibnamefont {Englert}}, \bibinfo {author} {\bibfnamefont {L.~L.}\
  \bibnamefont {S{\'{a}}nchez-Soto}}, \ and\ \bibinfo {author} {\bibfnamefont
  {E.}\ \bibnamefont {Karimi}},\ }\bibfield  {title} {\textit {\bibinfo
  {title} {Experimental investigation of high-dimensional quantum key
  distribution protocols with twisted photons},}\ }\href {} {\bibfield  {journal} {\bibinfo  {journal}
  {{Quantum}}\ }\textbf {\bibinfo {volume} {2}},\ \bibinfo {pages} {111}
  (\bibinfo {year} {2018}{\natexlab{b}})}\BibitemShut {NoStop}%
\bibitem [{\citenamefont {Zhu}\ \emph {et~al.}(2021)\citenamefont {Zhu},
  \citenamefont {Tyler}, \citenamefont {Valencia}, \citenamefont {Malik},\ and\
  \citenamefont {Leach}}]{zhu2021high}%
  \BibitemOpen
  \bibfield  {author} {\bibinfo {author} {\bibfnamefont {F.}\ \bibnamefont
  {Zhu}}, \bibinfo {author} {\bibfnamefont {M.}\ \bibnamefont {Tyler}},
  \bibinfo {author} {\bibfnamefont {N.~H.}\ \bibnamefont {Valencia}},
  \bibinfo {author} {\bibfnamefont {M.}\ \bibnamefont {Malik}}, \ and\
  \bibinfo {author} {\bibfnamefont {J.}\ \bibnamefont {Leach}},\
  }\bibfield  {title} {\textit {\bibinfo {title} {Is high-dimensional photonic
  entanglement robust to noise?}}\ }\href@noop {} {\bibfield  {journal}
  {\bibinfo  {journal} {AVS Quantum Science}\ }\textbf {\bibinfo {volume}
  {3}},\ \bibinfo {pages} {011401} (\bibinfo {year} {2021})}\BibitemShut
  {NoStop}%
\bibitem [{\citenamefont {Bouchard}\ \emph {et~al.}(2021)\citenamefont
  {Bouchard}, \citenamefont {England}, \citenamefont {Bustard}, \citenamefont
  {Fenwick}, \citenamefont {Karimi}, \citenamefont {Heshami},\ and\
  \citenamefont {Sussman}}]{bouchard2021achieving}%
  \BibitemOpen
  \bibfield  {author} {\bibinfo {author} {\bibfnamefont {Fr{\'e}d{\'e}ric}\
  \bibnamefont {Bouchard}}, \bibinfo {author} {\bibfnamefont {Duncan}\
  \bibnamefont {England}}, \bibinfo {author} {\bibfnamefont {Philip~J}\
  \bibnamefont {Bustard}}, \bibinfo {author} {\bibfnamefont {Kate~L}\
  \bibnamefont {Fenwick}}, \bibinfo {author} {\bibfnamefont {Ebrahim}\
  \bibnamefont {Karimi}}, \bibinfo {author} {\bibfnamefont {Khabat}\
  \bibnamefont {Heshami}}, \ and\ \bibinfo {author} {\bibfnamefont {Benjamin}\
  \bibnamefont {Sussman}},\ }\bibfield  {title} {\enquote {\bibinfo {title}
  {Achieving ultimate noise tolerance in quantum communication},}\ }\href@noop
  {} {\bibfield  {journal} {\bibinfo  {journal} {Physical Review Applied}\
  }\textbf {\bibinfo {volume} {15}},\ \bibinfo {pages} {024027} (\bibinfo
  {year} {2021})}\BibitemShut {NoStop}%
\end{thebibliography}

\providecommand{\noopsort}[1]{}

\end{document}